\documentstyle[11pt,aaspp4]{article}

\newcommand{\unit}[1]{\ifmmode\,{\rm #1}\else$\,{\rm #1}$\fi}

\slugcomment{To appear in Astrophysical Journal, April 10, 1997}

\lefthead{Haarsma~{\it et~al.}}
\righthead{Radio Light Curves of Gravitational Lens B0957+561}

\begin{document}
\title{The 6~cm Light Curves of B0957+561, 1979-1994: \\
	New Features and Implications for the Time Delay}

\author{D. B. Haarsma\altaffilmark{1}, 
	J. N. Hewitt\altaffilmark{1}, 
	J. Leh\'ar\altaffilmark{2}, 
	and B. F. Burke\altaffilmark{1}}

\altaffiltext{1}{Department of Physics, 26-335, Massachusetts 
Institute of Technology, Cambridge, MA 02139}

\altaffiltext{2}{Harvard-Smithsonian Center for Astrophysics, 
60 Garden Street, Cambridge, MA 02138}


\begin{abstract}

We report on 15~years of VLA monitoring of the gravitational lens
B0957+561 at 6~cm.  Since our last report in 1992, there have been 32
additional observations, in which both images have returned to their
quiescent flux density levels and the A image has brightened again.
We estimate the time delay from the light curves using three different
techniques: the $\chi^2$ analysis of Press, Rybicki, \&~Hewitt
(1992a,b), the dispersion analysis of Pelt~{\it et~al.} (1994, 1996),
and the locally normalized discrete correlation function of
Leh\'ar~{\it et~al.}  (1992).  Confidence intervals for these time
delay estimates are found using Monte Carlo techniques.  With the
addition of the new observations, it has become obvious that five
observations from Spring~1990 are not consistent with the statistical
properties of the rest of the light curves, so we analyze the light
curves with those points removed, as well as the complete light
curves.  The three statistical techniques applied to the two versions
of the data set result in time delay values in the range 398 to
461~days (or 1.09 to 1.26~years, A leading B), each with $\sim$5\%
formal uncertainty.  The corresponding flux ratios (B/A) are in the
range 0.698 to 0.704.  Thus, the new features in the light curve show
that the time delay is less than 500~days, in contrast with analysis
of earlier versions of the radio light curves.  The large range in the
time delay estimates is primarily due to unfortunate coincidences of
observing gaps with flux variations.

\end{abstract}



\section{Introduction}
\label{intro}

With the discovery of the double quasar B0957+561 in 1979
(\cite{walsh79}), the door was opened for observational cosmology
using gravitational lenses.  Lenses offer the possibility of
determining angular diameter distances at large redshifts, independent
of the assumptions that underlie other distance measurement
techniques.  The time delay between pairs of lensed images of a
variable source, when combined with a model of the lensing potential,
gives an estimate of angular diameter distance, and thus constrains
the Hubble parameter $H_0$, the deceleration $q_0$, and the
cosmological constant $\Lambda_o$ (Refsdal 1964, 1966;
\cite{narayan91}).  Since cosmological parameters other than 
$H_0$ affect the time delay, measurements of many lensed systems at
various redshifts will ultimately be needed.  The possibilities for
B0957+561 were realized immediately, and VLA and optical monitoring
for the time delay began in 1979.

Table~1 lists measurements of the time delay of B0957+561, showing the
literature reference, light curves, and estimate of the time delay in
days and in years.  Note that before 1989 the results were scattered
in delay, with large errors.  More recently, estimates for the delay
have clustered around 415~days and 540~days.  Before 1992 it seemed
that the optical light curves gave shorter delays and the radio light
curves gave longer delays, and since 1992 a variety of analyses have
been used on the light curves in attempts to resolve the
discrepancy. Although some statistical techniques have found
consistent delays at optical and radio wavelengths, the techniques
have not been consistent with each other.  The statistical properties
of the noisy irregularly sampled light curves (complicated by
microlensing effects at optical wavelengths) have proven difficult to
understand, and the conclusion has often been that more observations
are needed.  We present here lengthened radio light curves with
additional features.  

Although the ultimate goal of this project is to determine the angular
diameter distance to the lens, we focus this paper primarily on the
value of the time delay and techniques for measuring it accurately.
Determining $H_0$ from the time delay requires assumptions about $q_0$
and $\Lambda_0$, and a good model of the lensing potential, in itself
a subject of much recent work (Grogin \&~Narayan 1996a, 1996b;
\cite{bernstein93}; \cite{falco91a}; \cite{kochanek91}).  Recent 
observations have provided an improved understanding of the cluster
and galaxy lensing potentials (\cite{garrett94}; \cite{angonin94};
\cite{dahle94}; \cite{fischer96}; \cite{falco96}), which should allow 
for better constrained models.

In addition to the time delay, the light curves give the flux ratio
between the A and B images, an important parameter in lens modeling.
Since the lensing effect is achromatic, this relative magnification is
independent of wavelength, as long as the emission region size is also
independent of wavelength and the angular resolution of the
observations is comparable.  These conditions are not met when
comparing the VLA flux ratio to the optical flux ratio, since the
optical emission region is much smaller than the 6~cm emission region.
The VLA flux ratio will also differ from the VLBI ratio, since the VLA
beam averages over the varying magnification on mas scales
(\cite{conner92}).  All flux ratios reported here are ratios between
the VLA 6~cm images of A and B.

It is four years since our last report on the VLA 6~cm monitoring of
B0957+561.  Preliminary results on the most recent data have been
released (\cite{melbourne}; \cite{aasjan96}); here we provide the
complete data and results of 15~years of observations.  New features
have appeared that allow us to improve our estimate of the time delay.
To determine the time delay, we have chosen three of the techniques
referred to in Table~1: PRH$\chi^2$ analysis (Press, Rybicki,~\&~Hewitt,
1992a,b, hereafter PRHa, PRHb; \cite{rybicki92}; \cite{rybicki94}),
dispersion analysis (Pelt~{\it et~al.} 1994, Pelt~{\it et~al.} 1996,
hereafter P94, P96), and discrete correlation function analysis
(Leh\'ar~{\it et~al.} 1992, hereafter L92; \cite{edelson88}).  These
were selected to provide consistency with our previous work and to
explore an additional technique.  All three avoid interpolation of the
light curve, a practice that biases the result for irregularly sampled
data by weighting assumed data during observing gaps the same as real
observations (Falco~{\it et~al.} 1991b; PRHa; L92).  The application
of the selected methods to the same light curves, described in similar
notation and tested with the same Monte Carlo data, may help to
clarify some of the issues related to this problem.

In Section~2 we report our observations and the recent features in the
light curve.  Section~3 describes the synthetic data used to determine
the confidence interval for each statistic.  In Section~4 we use 
PRH$\chi^2$ analysis to find the time delay, and discuss issues
related to this method.  In Section~5 we report the results of
dispersion analysis of these light curves, and in Section~6 we discuss
the results of the discrete correlation function analysis.  Our
conclusions are in Section~7.


\section{Observations}
\label{obs}

The gravitational lens B0957+561 has been observed about once a month
from 1979 to the present at the National Radio Astronomy Observatory
(NRAO) Very Large Array radio telescope (VLA)\footnote{The National
Radio Astronomy Observatory is operated by Associated Universities,
Inc., under cooperative agreement with the National Science
Foundation.}.  Results through 1990 were reported by L92.  Since then,
32 observations have been added to the light curves, yielding 112
observations of good quality between 1979~June and 1994~December.  The
observations are made at two bands, 4.885 and 4.835~GHz (both
$\sim6\unit{cm}$), with 50~MHz bandwidth each.  Here we report only
the 4.885~GHz observations since the other band was not in use in the
early 1980s at the VLA.

The VLA cycles through four configurations (A, B, C, and D) about once
every 480~days.  At 6~cm, the angular resolution of the four arrays is
approximately 0\arcsec.3, 1\arcsec, 3\arcsec, and 10\arcsec,
respectively.  Since the separation of the A and B images in B0957+561
is 6\arcsec, the images are not resolved in the D configuration,
causing gaps in the monitoring of approximately 120~days in every
480~day period.  Of the 112 observations, 50 are from A, 31 from B,
and 31 from C.

All of the observations have been reduced using the techniques of L92,
in order to keep the treatment of the data as uniform as possible over
the 15~years.  All were flux- and phase-calibrated to the nearby point
source 1031+567, which was found by L92 to not vary by more than 2\%
in flux density with respect to the VLA flux calibrator 3C286.  After two
iterations of phase self-calibration, the data were cross-calibrated
to a reference map to align the coordinate systems for all the maps.
The reference map for each VLA configuration was made from the
observation with the best UV-coverage for that array, and we have used
the same reference maps as L92.  Finally, the extended structure in
the map was subtracted from each data set, leaving behind only the two
point images of the quasar core.  A Gaussian was fitted to each image
to determine its flux density.

The final light curves are reported in Table~2.  Although the flux
densities are reported in mJy, all of the real and synthetic data in
this paper were converted to ``dBJ'' units for analysis.  For a flux
density $S$ the conversion to dBJ is
\begin{equation}
	S\unit{dBJ} = 10 \log \left( \frac{S}{1\unit{Jy}}\right). 
\end{equation}
This logarithmic scale is similar to the optical magnitude scale, and
we have used it in order to be consistent with PRHb.   In these
units, a 2\% change in $S$ is 0.088\unit{dBJ}.  

Due to the deconvolution and self-calibration techniques involved in
VLA data reduction, the accuracy of the flux densities listed in
Table~2 cannot be determined analytically.  L92 estimated the errors
on these measurements in three ways: as the RMS during the quiescent
period (1983.3 to 1988.0 for A and 1984.5 to 1989.5 for B), as the RMS
in the residuals to a 2nd-order polynomial fit to the long decline
(1980.5 to 1984.0), and by splitting a single observation into
subsections in time.  L92 concluded that the errors are approximately
2\% of the flux density, which is about 0.6~mJy for the A image, and
about 0.4~mJy for the B image.  Due to the different synthesized beams
in the three VLA arrays used (A, B, C), it is possible that the errors
are significantly different for these arrays. Table~2 lists the VLA
array at the time of observation (sometimes a hybrid array), and the
array assumed during the data reduction (one of the standard arrays,
either A, B, or C, whichever is most similar to the observation
array).  For simplicity we have assumed a homogeneous error model for
each light curve, such that every data point has the same fractional
error, irrespective of the array of observation.

The 6~cm VLA light curves are plotted in Figure~\ref{lightcurve}.
Since 1990, the A and B images have both declined to approximately the
flux density of the mid 1980s.  In addition, the A image rose in flux
density during 1993.  We have been closely monitoring the B image in
1995-96, which has also increased in flux density, and these data will
be reported in a future publication.  This continued variation is
fortunate for those interested in measuring angular diameter distances
through the B0957+561 time delay.  As more features enter the curve,
the determination of the delay should continue to improve.  Note that
the two light curves have very similar features, with no significant
differences on long time scales.  This is evidence that the radio
light curves, unlike the optical light curves, are not affected by
microlensing on time scales of one to fifteen years.

In Spring~1990, the B image changed in flux density by nearly 4~mJy in
a few months.  These data points have already been the subject of
considerable discussion (\cite{kayser93}; P94; P96).  We have looked
carefully at the raw data, and find that there were no abnormalities
in these observations: there were no weather problems, no bad
antennas, and the self-calibration, mapping, and subtraction of
extended emission all proceeded smoothly. The final maps had no
artifacts from the reduction and were of low noise.  The flux
densities of A and B observed with the second VLA band (4.835~GHz)
were slightly different than the reported data (about 0.15\unit{mJy}
for these data sets), a difference which is typical for data sets at
other times in the light curve (\cite{sopata95}).  This implies that
there was no frequency-dependent measurement errors or corruption of
the data.  Thus, we conclude that the fluctuation of Spring~1990 is of
physical origin and that the points cannot be excluded as poor quality
data.  Possible explanations for this fluctuation might be refractive
interstellar scintillation (RISS), microlensing in the lensing galaxy,
or intrinsic variation of the quasar (if the variation in the A image
occurred in a gap, such as Summer~1988).  All of these processes would
require the source to be extremely small to achieve a fluctuation time
scale of a few months.  Since the fluctuation occurs during the start
of an outburst, it is possible that the emission is from a compact jet
component emerging from the core.  The feature in 1989-91 (A image)
and 1990-92 (B image) would be due to the component expanding and then
cooling.  Such a component could be small enough to be susceptible to
RISS or microlensing.


\section{Synthetic Data}
\label{monte}

To find the uncertainty in the time delay estimates we made a suite
of synthetic data sets.  We will describe these synthetic data here,
then refer to them as they are used in later sections.

We made a set of 500 Gaussian Monte Carlo light curve pairs, with 112
points in each curve at the same observation times as the real radio
light curves.  We will use the notation N=x to denote different
lengths of the light curves, where x is the number of points in the
curve.  The curves were generated using a Gaussian process, with
the same structure function as that assumed for the real light curves
(see Section~4, eq.~[\ref{N=107sf.eq}]).  The measurement errors were
modeled as Gaussian random variables with zero mean and RMS
$e_A=0.073$~dBJ (image A) and $e_B=0.087$~dBJ (image B)
(eq.~[\ref{N=107err.eq}]).  The data were then given a randomly chosen
time delay and flux ratio, uniformly distributed in the ranges 350 to
650~days in delay and 0.68 to 0.72 in ratio.

In each of the following sections, these 500 Gaussian Monte Carlo data
sets are used to determine the uncertainty in the time delay and flux
ratio values found for the real N=112 light curves.  This is done by
determining the minimum with respect to delay and ratio of the
PRH$\chi^2$ and dispersion surfaces, or the maximum with respect to
delay of the discrete correlation function, for each of the 500 data
sets.  The differences between the fitted and true delays (and the
fitted and true flux ratios) are then found for all the Monte Carlo
data.  The median of the fitted-minus-true values measures the bias in
the result, and we subtract this value from the fitted delay and ratio of
the real data.  To find the 68\% confidence interval, we count out 170
data sets on both sides of the median, enclosing 68\% of the points,
then adjust the interval for the bias by subtracting the median.

We also did a quasi-jackknife analysis of the data.  P94 chose to
remove some points from the radio light curve, choosing the points
that seemed to have the biggest influence on the time delay.  To do a
more formal analysis of the effect of leaving out points, we made a
second set of synthetic data by leaving out each data point, one at a
time.  This created 112 data sets, each with 111 data points.  This is
the procedure used when doing a ``jackknife'' analysis on data that is
 {\em not} in a time series.  It is not formally correct for a time
series, because the data are correlated in time and do not meet the
necessary condition of being independent and identically distributed.
Our purpose in using these ``pseudo-jackknife'' data sets is to
demonstrate the impact of leaving out individual points, and to create
data sets with statistical properties identical to the real data,
even if those properties are non-Gaussian or non-stationary.

As mentioned in Section~2, the B image has a strong variation in
Spring~1990 which is not due to measurement error and must be of
physical origin.  For reasons explained in Section~4, we have chosen
to do our analysis for both the complete light curves and for the
light curves with five data sets from Spring~1990 removed, leaving 107
data points.  In order to find the confidence intervals in the N=107
analysis, we made 500 Gaussian process data sets in the same manner as
described above, except that each contains 107 data points.  We have
also made 107 pseudo-jackknife data sets of 106 points each.

Thus, we have four sets of synthetic data: 500 Gaussian process Monte
Carlo data sets with 112 points each, 500 Gaussian process Monte Carlo
data sets with 107 points each, 112 pseudo-jackknife data sets of 111
points each, and 107 pseudo-jackknife data sets of 106 points each.


\section{PRH$\chi^2$ Analysis}
\label{prhchi}

\subsection{Technique} 

Press, Rybicki, and Hewitt (PRHa, PRHb) have determined time delays from
the published optical and radio light curves (Vanderriest~{\it et~al.}
1989; L92) using structure function analysis and a chi-squared fitting
technique.  Rybicki~\&~Press (1992) and Rybicki~\&~Kleyna (1994)
further describe the technique and present some modifications.  In
this section we apply the technique to the new radio light curves.

Following PRHa, we describe the measured light curve $y(t)$ as the sum
of the signal $s(t)$ from the source and the noise signal $n(t)$
representing the measurement error 
\begin{equation}
	y(t) = s(t) + n(t).
\end{equation}
Given the covariance matrix associated with $s(t)$, $C_{ij} = \langle
s(t_i)s(t_j) \rangle$, the covariance model {\bf B} associated with
$y(t)$ has elements 
\begin{equation}
	B_{ij} = C_{ij} + \langle n_i^2 \rangle \delta_{ij},
\end{equation}
and its inverse is $ {\bf A} = {\bf B}^{-1}$.  The joint probability
distribution of the data vector {\bf y}, assuming a Gaussian process,
is
\begin{equation}
\label{prob.eq}
	P({\bf y}) = [(2 \pi)^N \det {\bf B}]^{-1/2} \exp \left[ - {1 \over 2} \chi^2 \right ].
\end{equation}
The PRH technique is to minimize $\chi^2$, 
\begin{equation}
	\chi^2 = \sum_{ij} (y_i - \bar{y}) A_{ij}(y_j - \bar{y})
\end{equation}
where 
\begin{equation}
	\bar{y} = \frac {\sum_{ij} y_i A_{ij} }{ \sum_{ij} A_{ij} }
\end{equation}
is the estimate of the mean of $y_i$ obtained by minimizing $\chi^2$
with respect to the unknown value of $\bar{y}$ (one degree of freedom
is lost in this estimate).

The PRH$\chi^2$ statistic is a measure of the goodness-of-fit of a
light curve to a model that assumes the temporal correlations of the
light curve are described by the covariance matrix $B_{ij}$.  It was
shown by PRHa to be independent of the underlying mean and variance of
the time series.  PRHa,b estimated the time delay by adopting trial
delays $\tau$ and trial flux ratios $R$, combining the light curves
according to these trial values, and computing the PRH$\chi^2$
statistic of the combined light curve for each ($\tau$, $R$) pair
using the covariance matrix that describes a single light curve.  Note
that the PRH$\chi^2$ sum includes contributions from all pairs of
points $y_i, y_j$ in the light curve under consideration ($(2N)^2$ pairs
for the combined curve).  Each point is compared to every other point
to test how well that pair fits the model, and thus all information
available in the light curve is used.

Rybicki~\&~Kleyna (1994) point out that {\it both} the PRH$\chi^2$
statistic and the normalization factor in the joint probability
distribution (eq.~[\ref{prob.eq}]) are functions of parameters of the
correlation model.  Therefore, minimizing only the PRH$\chi^2$
statistic does not give a true measure of the most likely dataset.
Rather, maximum likelihood estimators are found by minimizing the
following quantity:
\begin{equation}
	Q = \log \det {\bf B} + \sum_{ij} (y_i-\bar{y}) A_{ij} (y_j-\bar{y}).
\end{equation}
Since the time delay is a parameter of the covariance model that
applies to the combined light curves, the neglect of $\log\det {\bf B}$
in the PRH$\chi^2$ minimization procedure is a concern.  The effect of
neglecting the $\log\det {\bf B}$ term is to favor samplings of the
combined light curve that discriminate less among different time
delays.  In other words, time delays are favored in which the overlap
is small when the light curves are combined.  Thus, the use of $Q$
rather than PRH$\chi^2$ will, in part, alleviate the impact of
sampling on the result.  For this data set, however, we find that the
value of $\log\det {\bf B}$ does not change by more than a few (in
units of $\chi^2$) for delays in the range 200--800~days.  Also, the
location of the minimum in delay-ratio space for the $Q$ and
PRH$\chi^2$ surfaces does not differ by more than two days in delay.
We also found that for ``window function'' data (light curves of
constant flux density but the same sampling as the real light curves),
$Q$ was not immune to the sampling and varied over the range of delays
in a way similar to PRH$\chi^2$.  Since it is the correct maximum
likelihood estimator in this problem, we report the time delay that
minimizes the $Q$ statistic, although it differs only slightly from
the time delay that minimizes PRH$\chi^2$.

In order to use $Q$ or PRH$\chi^2$, the covariance model $B_{ij}$ for
the data must be determined.  The variability properties of the light
curves can be described by a first-order structure function
(\cite{simonetti85})
\begin{equation}
	V(T) = {1 \over 2} \langle [s(t) - s(t-T)]^2\rangle,
\end{equation}
where $T$ is the lag between two points on the light curve.  We assume
a power law form for this structure function, which is typical of
quasar light curves (\cite{hughes92}), and make a fit to the power law
using point estimates found directly from the data
\begin{equation}
	v_{ij} = \frac{1}{2}\left[ (y_i - y_j)^2 - e_i^2 - e_j^2 \right],
\end{equation}
where $e_i$ is the measurement error for $y_i$.  If we assume the
light curves are stationary, then $C_{ij} = C(t_i - t_j) = C(T)$, and we
can relate this autocorrelation function to the structure function by
\begin{equation}
	C(T) = \langle s^2\rangle - V(T), 
\end{equation}
where $\langle s^2 \rangle$ is the variance of the signal $s(t)$.
Thus, from an expression for $V(T)$, elements of the covariance matrix
are computed by
\begin{equation}
	B_{ij} = B(t_i - t_j)= \langle s^2\rangle - V(t_i - t_j) + e_i^2\delta_{ij}.
\end{equation}
As shown by PRHa, the PRH$\chi^2$ and $Q$ statistics are independent
of the assumed value of the variance $\langle s^2\rangle$.

An optimal reconstruction $\hat{s}(t)$ of the underlying signal $s(t)$
can be found by minimizing the squared difference $\langle [\hat{s}(t)
- s(t)]^2\rangle$ between them for each time.  PRHa show that
\begin{equation}
	\hat{s}(t) = \sum_{ij} \langle s(t_j)s(t)\rangle A_{ij} (y_i - \bar{y}),
\end{equation}
and thus the optimal reconstruction $\hat{s}(t)$ can be found from
{\bf y} once $A_{ij}$ is known.

\subsection{PRHb Covariance Model} 

\begin{figure}
\plotfiddle{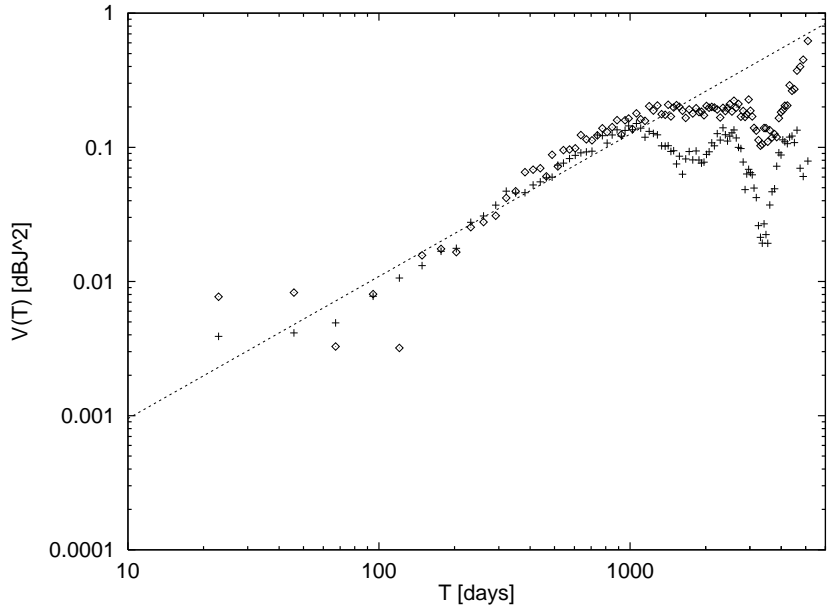}{3in}{-90}{45}{45}{-200}{250}
\caption{Point estimates of the structure function for the N=112 light 
curves.  The crosses are from the A light curve, and the diamonds are
from the B light curve.  The dotted line is the structure function
found by PRHb.}
\label{N=112sf.prh}
\end{figure}

We have applied the above analysis to the N=112 light curves, following 
exactly the PRHb analysis of the N=80 light curves.  We assumed the
same power-law structure function,
\begin{equation}
\label{N=112sf.eq}
V(T) = 8.28\times10^{-5}\unit{dBJ^2} \left(\frac{T}{1\unit{day}}\right)^{1.06},
\end{equation}
and the same error estimate,
\begin{eqnarray}
\label{N=112err.eq}
e_A = 0.047\unit{dBJ} = 1.1\%, & e_B = 0.088\unit{dBJ} = 2.0\%.
\end{eqnarray}  
This fit of the structure function to the point estimates $v_{ij}$ of
$V(T)$ is shown in Figure~\ref{N=112sf.prh} for N=112.

\begin{figure}
\plotfiddle{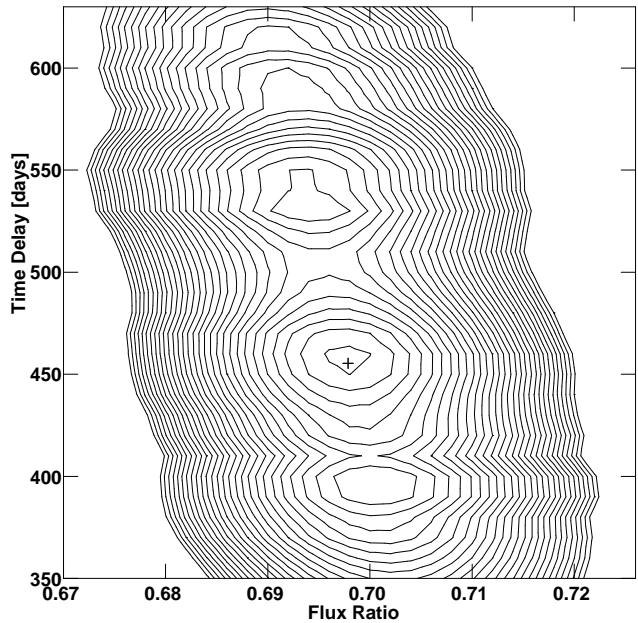}{3in}{0}{45}{45}{-135}{-75}
\caption{The PRH$\chi^2$ surface for N=112 data points, using the PRHb
covariance model.  The global minimum is PRH$\chi^2 = 283$ at $\tau =
455$~days, R = 0.6979, for 221 degrees of freedom.  Contours start at
PRH$\chi^2 = 285$ and increase by 5 to 430.}
\label{N=112chi2surf}
\end{figure}

Using this model of the underlying quasar emission process and the
measurement error, we used a downhill simplex search, or ``amoeba''
(\cite{numrec}), to find the delay and ratio minimizing PRH$\chi^2$.
The PRH$\chi^2$ surface is plotted in Figure~\ref{N=112chi2surf}.  For
221 degrees of freedom (2N minus three for $\tau$, $R$, and
$\bar{y}$), the global minimum is PRH$\chi^2 = 283$ at $\tau=455$~days
and $R=0.6979$.  This time delay is much shorter than the delay PRHb
found for the first 80 data points ($\tau = 548^{+19}_{-16}$~days,
95\% confidence interval).  The estimate of the delay has changed
significantly with the addition of new features to the light curves.

There are, however, several reasons to suspect this result for the
N=112 data set.  First, for 221 degrees of freedom and Gaussian light
curves, $\chi^2 \geq 283$ has a probability of 0.3\%.  If the real
light curves are indeed a Gaussian process with Gaussian noise, we can
use PRH$\chi^2$ as a measure of goodness-of-fit.  Even if not,
PRH$\chi^2$ of the A light curve alone plus PRH$\chi^2$ of the B light
curve alone should equal PRH$\chi^2$ of the combined curve, if the
covariance model, delay, and ratio are correct.  Here, PRH$\chi^2$=111
(for A) plus PRH$\chi^2$=116 (for B) equals 227, which is much less
than 283, indicating a bad fit.  For N=80, PRH$\chi^2$ was also large,
but the probability (8\%) was more acceptable (PRHb).

Second, the PRH$\chi^2$ surface has several secondary minima,
including one at $\tau \simeq 530$~days with PRH$\chi^2 \simeq
295$. While the minimum at 455~days is formally more significant than
the minimum at 530~days, it still raises some doubt about which is the
best delay for the data set.

\begin{figure}
\plotfiddle{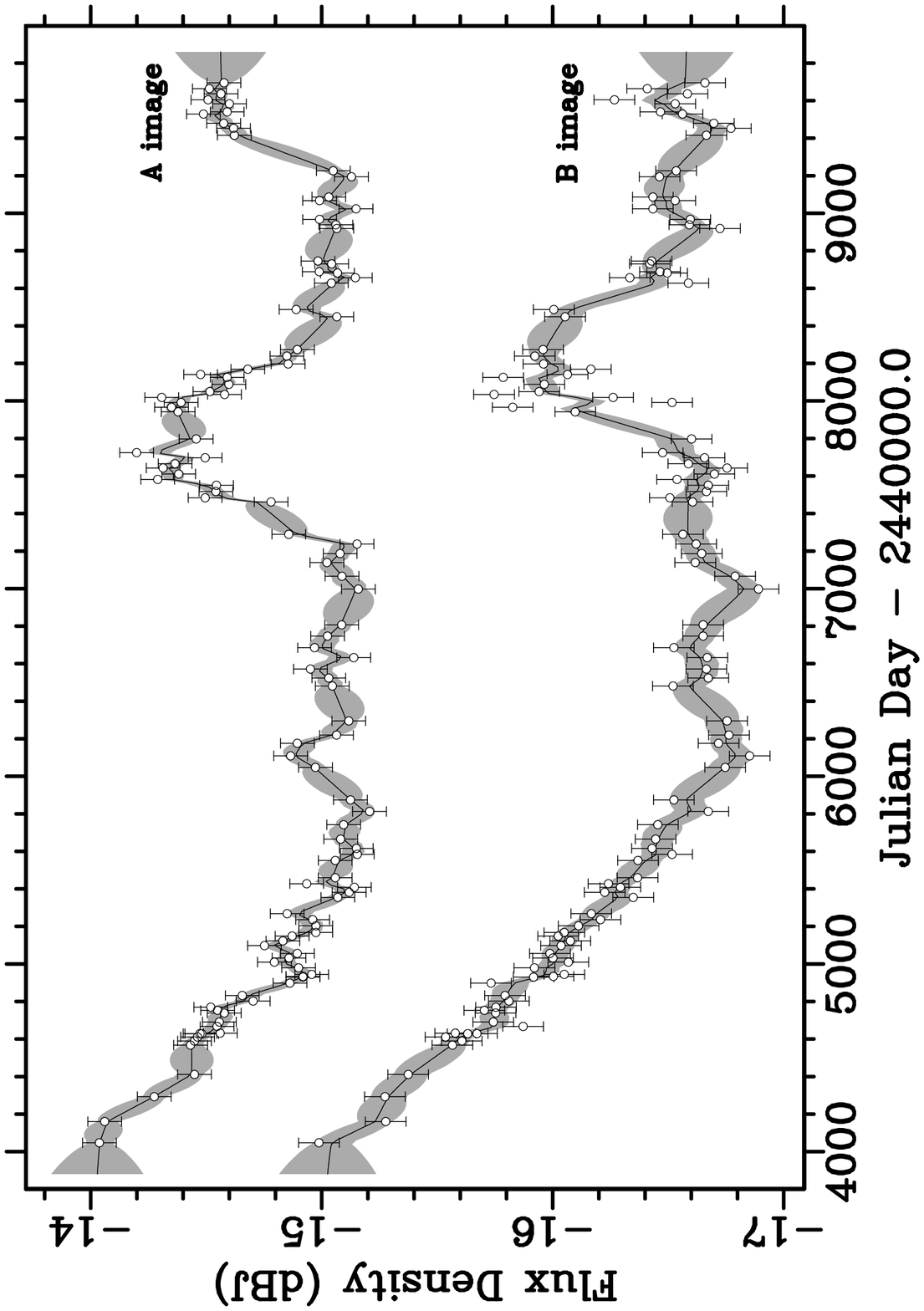}{3in}{-90}{50}{50}{-200}{275}
\caption{Optimal reconstructions for the N=112 light curves, using 
the PRHb covariance model.  The gray region is the one $\sigma$ error
about the reconstruction.  }
\label{N=112recon}
\end{figure}

\begin{figure}
\plotfiddle{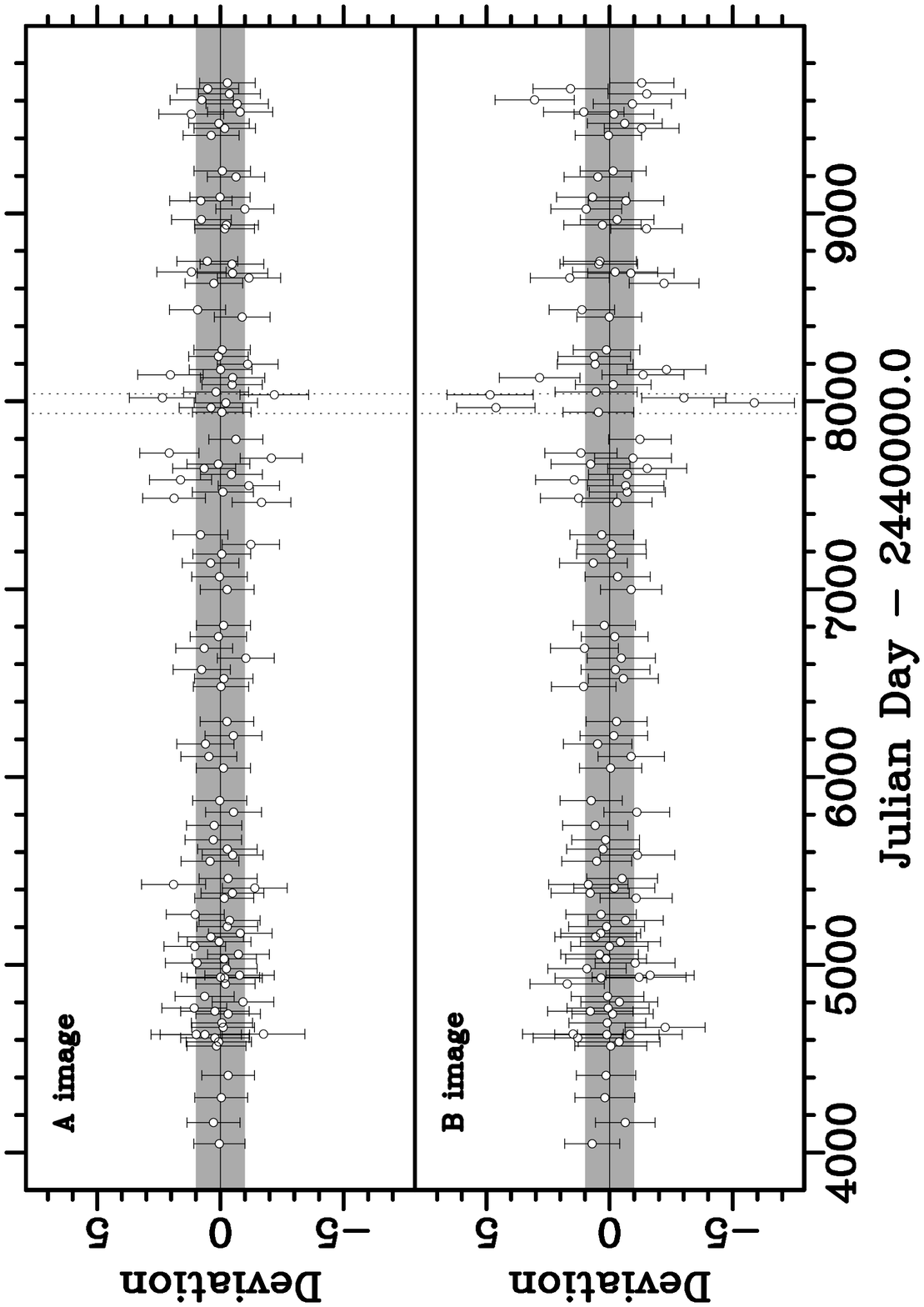}{3in}{-90}{50}{50}{-200}{275}
\caption{Differences between the real data and the optimal reconstruction for 
the N=112 light curves, using the PRHb covariance model.  The gray
region is the one $\sigma$ error about the reconstruction.  The
observed data points and their errors were normalized so that the one
$\sigma$ band about the optimal reconstruction was unity.  The epoch
removed is marked with dashed lines. }
\label{N=112recondiff}
\end{figure}

Finally, the optimal reconstructions of the individual A and B N=112
light curves indicate problems (Figure~\ref{N=112recon}).  The
reconstruction of the A image light curve has a lot of short time
scale variation.  By eye, one would guess that many of the small
fluctuations in the real data are measurement error, but the
covariance model causes the reconstruction to interpret them as
signal.  Figure~\ref{N=112recondiff} shows the differences between the
real data and the optimal reconstruction.  For both light curves, we
have scaled the one $\sigma$ error bars of the reconstruction to equal
unity and applied this scale to the flux densities and errors of the
corresponding real observations.  For the B image, it can be seen that
several observations in Spring~1990 (around Julian day 2448000) have
an unusually high deviation from the optimal reconstruction.  We have
found no structure function for the individual light curves that
provides a better fit to the Spring~1990 points. Therefore, we believe
that they have {\em different} statistical properties than the rest of
the light curve.  This is evidence that a different (or additional)
physical process is at work during this epoch than during the rest of
the light curve.

The above concerns are indications that the assumed structure function
and measurement error (eqs.~[\ref{N=112sf.eq}] and
[\ref{N=112err.eq}]), found by PRHb for the N=80 data set, are not a
good covariance model for the N=112 data set; the additional
observations since 1990 show that a better covariance model for the
data is needed.  The optimal reconstructions show that the B image
data points from Spring~1990 cannot be fit with the same model as the
rest of the B image light curve.  Since we know these data are modeled
incorrectly in the individual curve, we remove them so that they will
not confuse the analysis of the combined curve.  Thus we repeat our
analysis with five consecutive observations from Spring~1990 removed:
1990~February~19, 1990~March~15, 1990~April~10, 1990~May~7, and
1990~May~23 (to be conservative, both the A and B measurements are
removed at each of these times).  All of the analysis in this paper
has been done with two versions of the light curve, N=112 points and
N=107 points.

\subsection{New Covariance Model} 

\begin{figure}
\plotfiddle{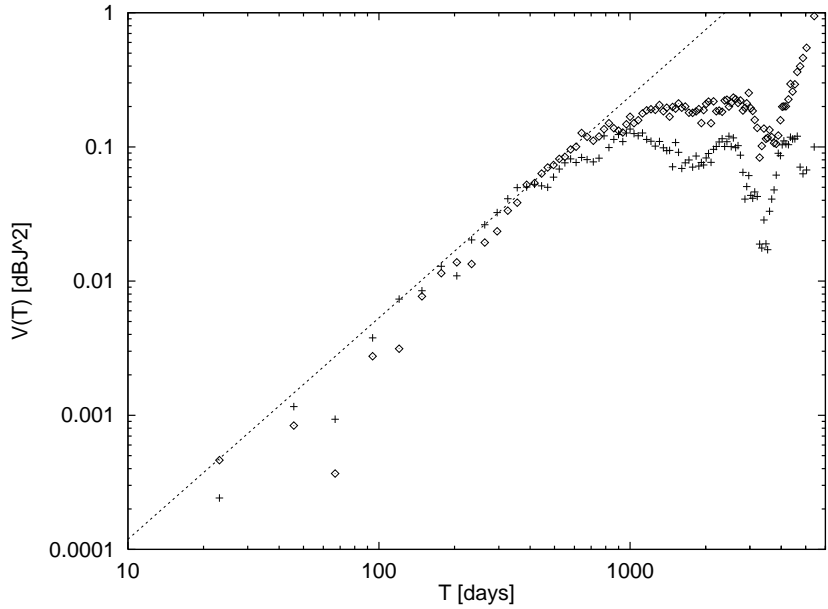}{3in}{-90}{45}{45}{-200}{250}
\caption{Point estimates of the structure function for the N=107 light 
curves.  The crosses are from the A light curve, and the diamonds are
from the B light curve.  The dotted line is our best fit for the
structure function.}
\label{N=107sf.best}
\end{figure}

\begin{figure}
\plotfiddle{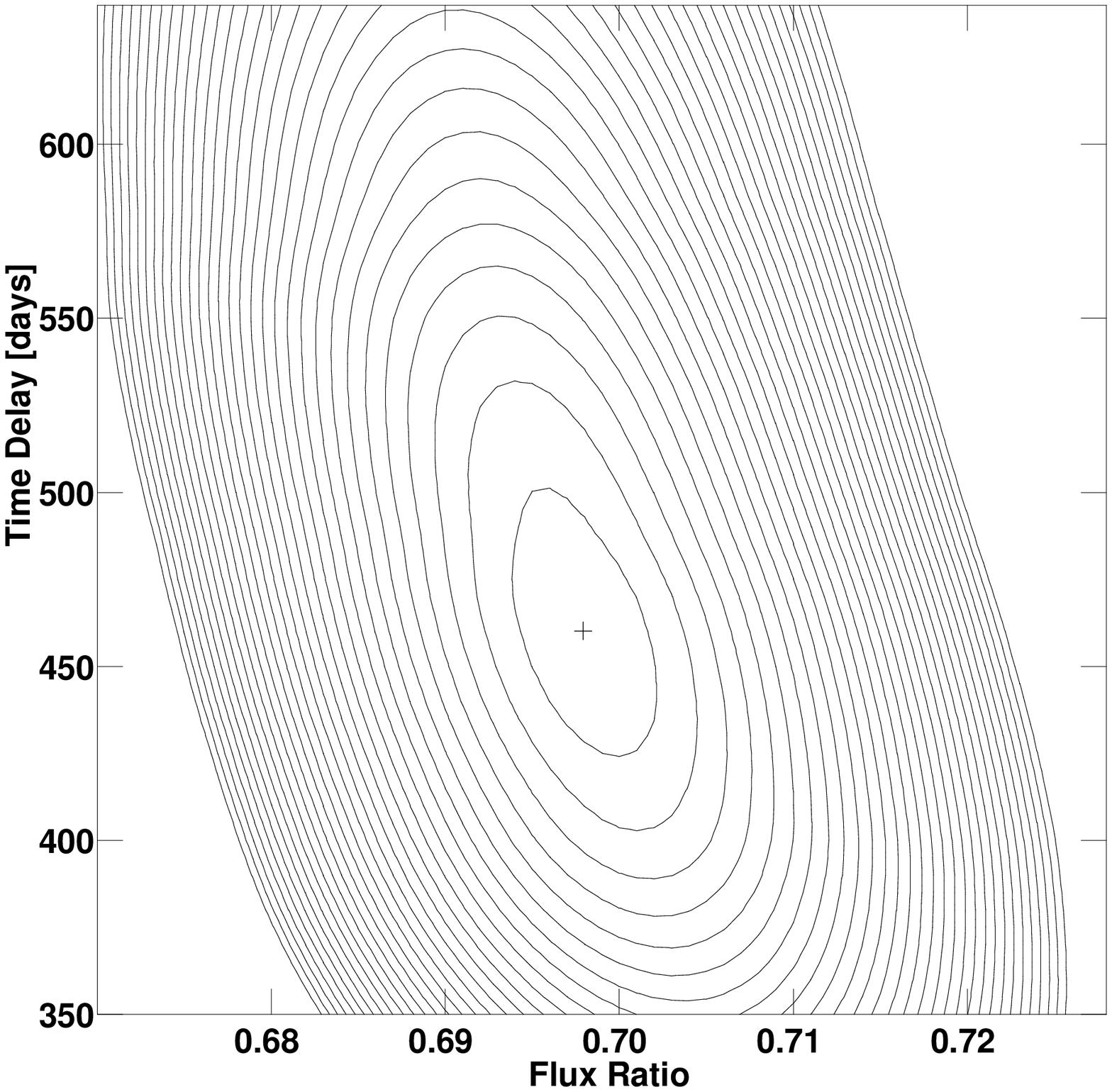}{3in}{0}{45}{45}{-135}{-75}
\caption{The $Q$ surface for N=107 data points, using the new
covariance model.  The global minimum is $Q=-719$ at $\tau = 460$~days
and $R = 0.6979$.  Contours start at $Q=-715$ and increase by five to
$Q=-570$.}
\label{N=107qsurf}
\end{figure}

With the removal of the Spring~1990 points, we believe the N=107 light
curves are a homogeneous data set for which a single structure
function accurately describes the underlying physical process for both
light curves.  To find the new covariance model, we use the PRH$Q$
statistic.  The $Q$ statistic allows for the parameters of the
covariance model to be fit at the same time as the delay and ratio, so
that the model is found directly from the {\em combined} data.  There
are seven parameters to fit: $\tau$, $R$, $e_A$, $e_B$, $\bar{y}$, and
the exponent and normalization of the power-law structure function.
We had difficulties with the minimization in seven-dimensional space,
so we instead found a preliminary fit for the measurement errors and
then used $Q$ to optimize the fit for the other parameters while
holding $e_A$ and $e_B$ fixed.  To find the measurement errors, we
made the point estimates to the structure function $v_{ij}$ for all
possible pairs in the individual N=107 light curves, assuming $e_A =
e_B = 0.080$~dBJ (about 2\% in flux density, as suggested by L92),
then fit a power law to the point estimates in the range
$200\unit{days} < T < 800\unit{days}$, and found a preliminary fit for
the structure function.  Using this preliminary fit, we adjusted $e_A$
and $e_B$ until PRH$\chi^2$ for each curve to the preliminary
structure function was approximately equal to the number of degrees of
freedom, obtaining
\begin{eqnarray} 
\label{N=107err.eq}
	e_A = 0.073\unit{dBJ} = 1.7\%, & e_B = 0.087\unit{dBJ} = 2.0\%.  
\end{eqnarray}
Next, we minimized $Q$ using a downhill simplex search (\cite{numrec})
for the combined light curve with respect to $\tau$, $R$, $\bar{y}$,
and the structure function parameters, while holding $e_A$ and $e_B$
fixed.  For the 209 degrees of freedom (2N minus five), the minimum
was $Q=-719$, with PRH$\chi^2=241$.  The best structure function fit
(see Figure~\ref{N=107sf.best}) was
\begin{equation}
\label{N=107sf.eq}
	V(T) = 2.664\times10^{-6}\unit{dBJ^2}\left(\frac{T}{1\unit{day}}\right)^{1.650}.
\end{equation}
Equations~\ref{N=107err.eq} and \ref{N=107sf.eq} are our best
covariance model for the light curves.  At the minimum, the time delay
and flux ratio are $\tau=460$ and $R=0.6979$.  The $Q$ surface in
delay and ratio is shown in Figure~\ref{N=107qsurf} (the PRH$\chi^2$
surface is very similar, with a minimum of 241 at $\tau=459$~days and
$R=0.6980$).  When the bias and 68\% confidence intervals are found
using the 500 N=107 Gaussian process Monte Carlo data sets, we obtain
\begin{eqnarray}
	\tau = 461^{+16}_{-15}, & R = 0.6981^{+0.0023}_{-0.0024} &
(N=107,\unit{PRH}Q).
\end{eqnarray}

The value of the delay for N=107 is very similar to that obtained for
the N=112 data set with PRH$\chi^2$.  This time, however, the result
is more reliable for several reasons.  First, the value of PRH$\chi^2$
at the $Q$ minimum is 241, and the probability that $\chi^2 \geq 241$
for 209 degrees of freedom is 6.4\%.  If the curves are Gaussian, we
can use PRH$\chi^2$ as a measure of goodness of fit, because $\log\det
{\bf B}$ is a constant (given a covariance model and sampling), and
thus $Q$ will have the same distribution as PRH$\chi^2$.  (The
PRH$\chi^2$ of the individual curves was already used to determine
$e_A$ and $e_B$, so it can no longer be used as a check on the
goodness of fit.)

Second, both the $Q$ and PRH$\chi^2$ surfaces for N=107 and the new
covariance model are smooth and have a single minimum, without the
secondary minima that characterized the N=112 analysis using the old
covariance model.  Note that the width of the global minimum has also
increased in delay space.

\begin{figure}
\plotfiddle{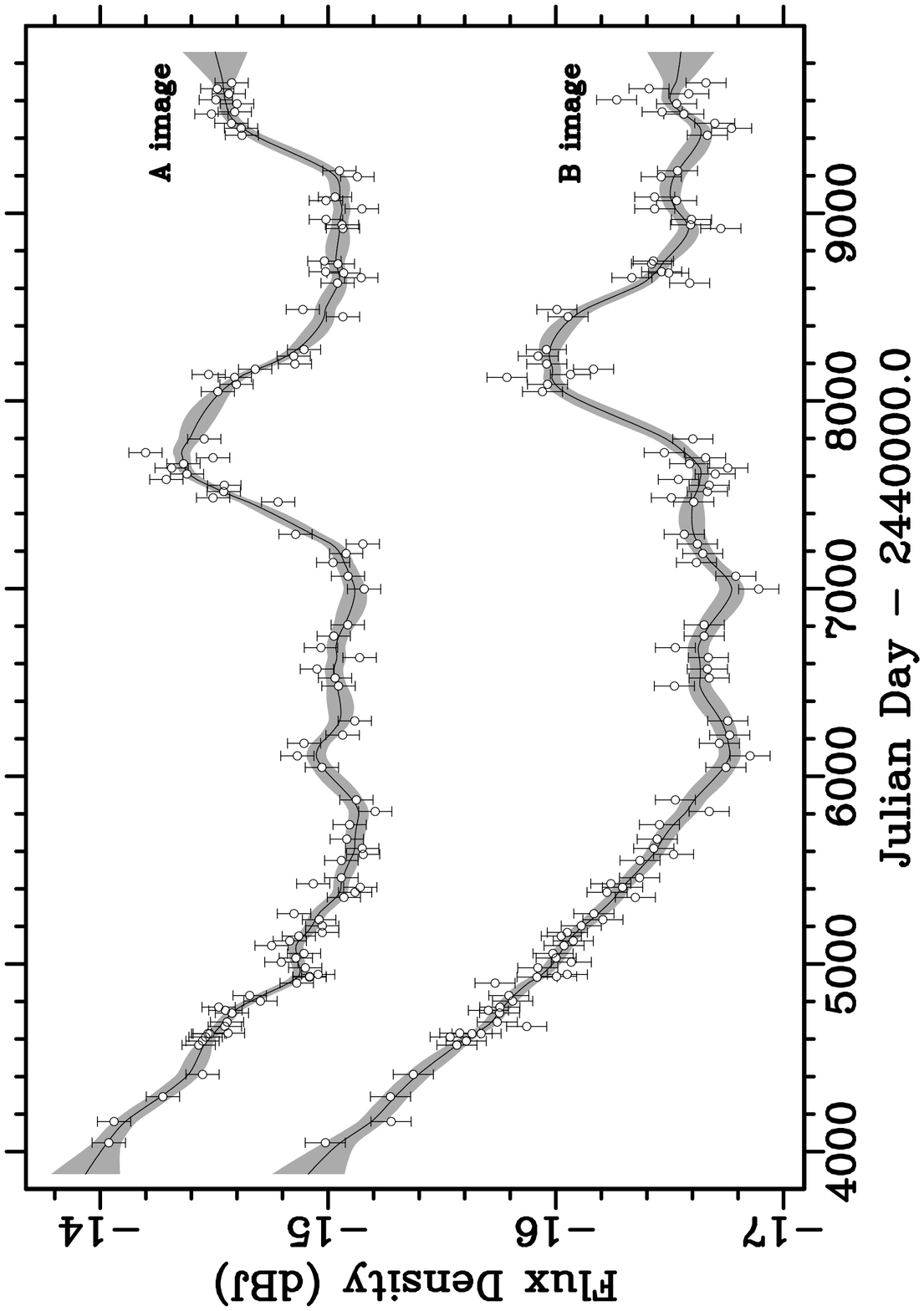}{3in}{-90}{50}{50}{-200}{275}
\caption{Optimal reconstructions for the N=107 light curves, using
the new covariance model.  The gray region is the one $\sigma$ error
about the reconstruction. }
\label{N=107recon}
\end{figure}

\begin{figure}
\plotfiddle{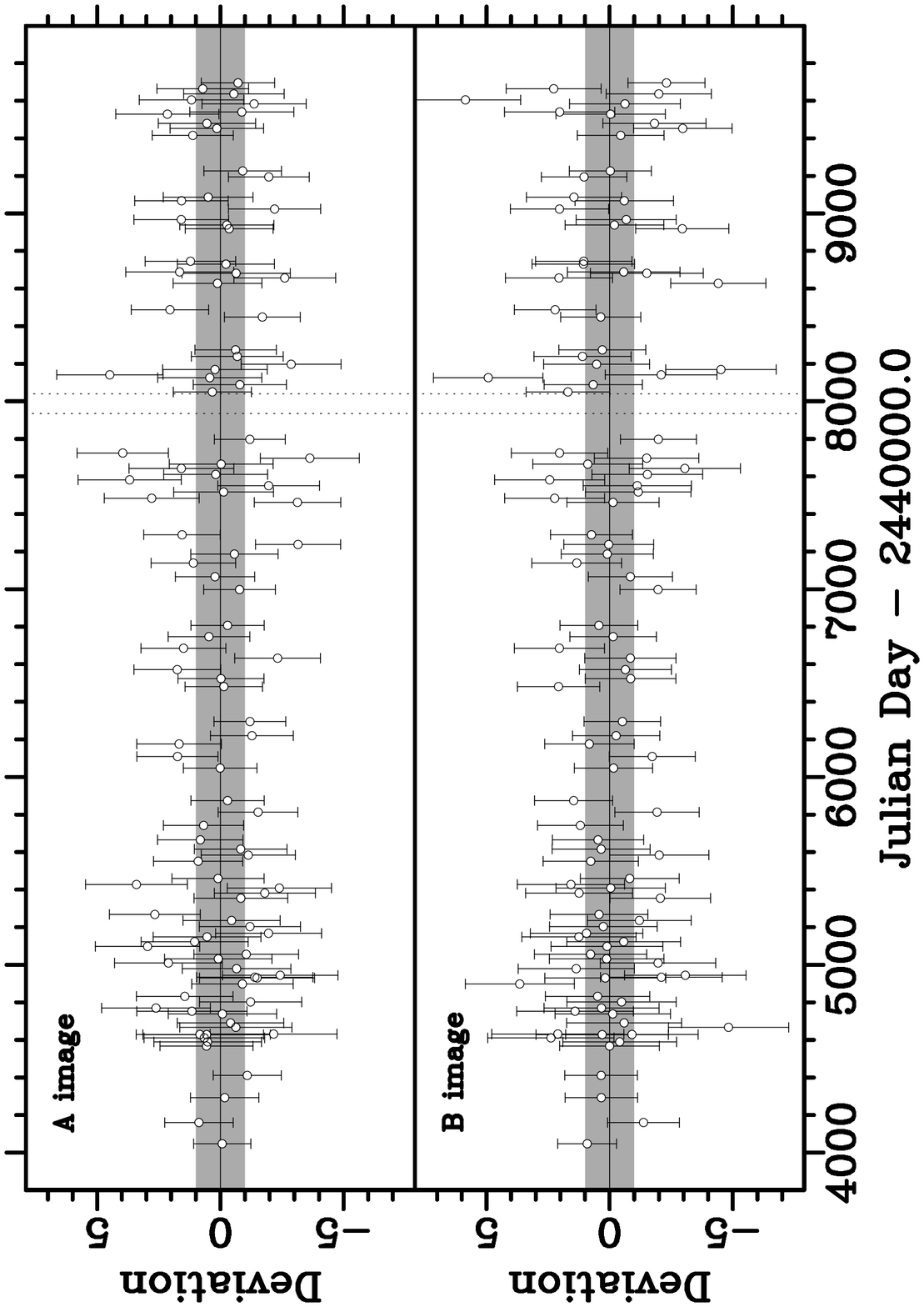}{3in}{-90}{50}{50}{-200}{275}
\caption{Differences between the real data and the optimal reconstruction for 
the N=107 light curves, using the new covariance model.  The gray
region is the one $\sigma$ error about the reconstruction.  The
observed data points and their errors were normalized so that the one
$\sigma$ band about the optimal reconstruction was unity. The epoch
removed is marked with dashed lines. }
\label{N=107recondiff}
\end{figure}

The optimal reconstructions of the individual N=107 curves are shown
in Figure~\ref{N=107recon}. The reconstruction of the N=107 A light
curve has less short time scale power than N=112, agreeing with our
guess by eye that most of the short time scale activity is measurement
error rather than signal.  The differences between the reconstruction
and the original data is shown in Figure~\ref{N=107recondiff}.  All of
the data now fall further from the reconstruction than before, but
there are no data points falling much further from the reconstruction
than their neighbors, as the Spring~1990 points did in
Figure~\ref{N=112recondiff}.

\begin{figure}
\plotfiddle{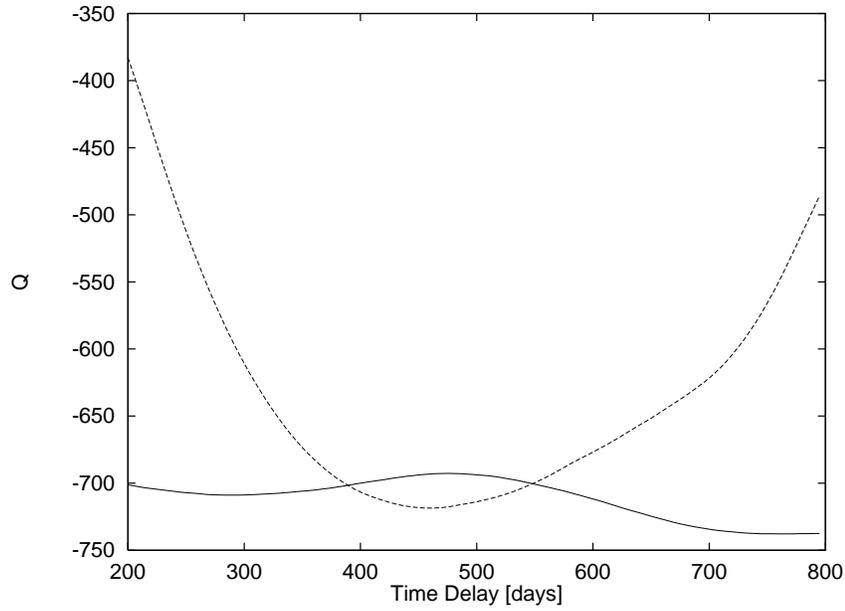}{3in}{-90}{45}{45}{-200}{250}
\caption{PRH$Q$ for the N=107 ersatz data 
(solid line), and for the real N=107 data (dotted line).  }
\label{N=107ersatz}
\end{figure}

To test the impact of sampling on our result, we made an ``ersatz'' or
``window function'' data set, where the light curves have the same
sampling as the real data but constant flux density.  The $Q$
statistic for the ersatz data is plotted in Figure~\ref{N=107ersatz},
along with the $Q$ statistic for the real data (where the flux ratio
has been set to $R=0.7$).  The ersatz data show a mild peak at about
480~days, corresponding to the VLA configuration cycle.  Delays of
480~days are excluded most strongly because at that delay there is the
most overlap between the A and B light curves.  Thus, the real data show a
minimum at a time delay of 460~days in {\em spite} of the sampling
effects.

\begin{figure}
\plotfiddle{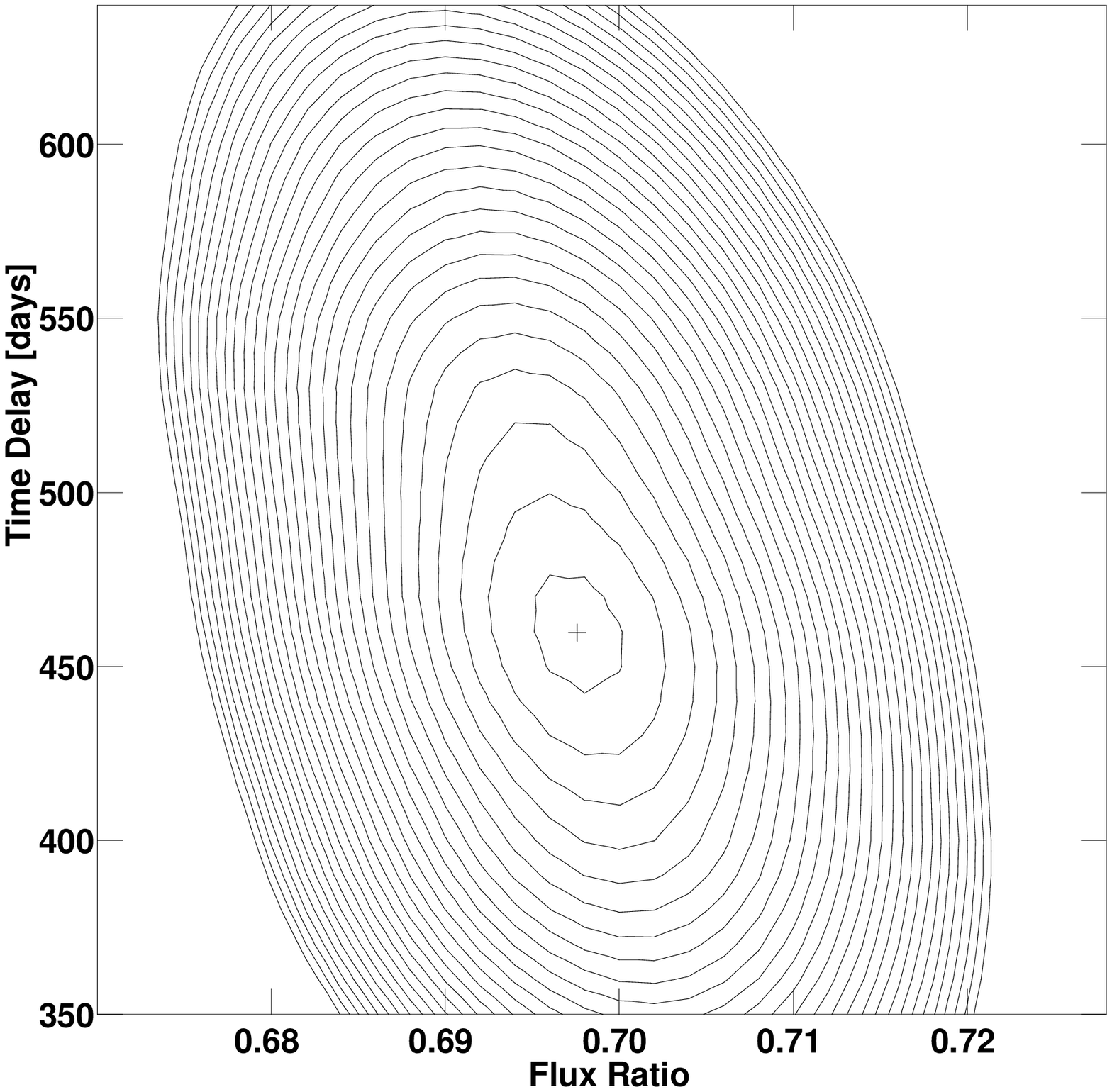}{3in}{0}{45}{45}{-135}{-75}
\caption{The $Q$ surface for N=112 data points, using the new
covariance model.  The global minimum is $Q=-707$ at $\tau = 460$~days
and $R = 0.6976$.  Contours start at $Q=-705$ and increase by five to
$Q=-570$.}
\label{N=112qsurf}
\end{figure}

With the covariance model for N=107 in hand, we can go back and use it
to analyze the N=112 light curve.  The N=112 $Q$ surface is shown in
Figure~\ref{N=112qsurf}, and we find that the surface is now smooth.
Thus, the change between Figures~\ref{N=112chi2surf} and
\ref{N=107qsurf} was due to the new covariance model, {\em not} to the
removal of points (or to the use of $Q$ instead of PRH$\chi^2$, since
$\log \det {\bf B}$ is nearly flat over the surface).  The minimum of
the N=112 surface is $Q=-707$ at $\tau=460$, $R=0.6976$, and
PRH$\chi^2=301$.  When the 500 N=112 Gaussian process Monte Carlo data
sets are analyzed with the same procedure, we find the bias and
confidence intervals and obtain
\begin{eqnarray}
	\tau = 459^{+14}_{-16}, & R = 0.6976^{+0.0022}_{-0.0023} & (N=112,\unit{PRH}Q).
\end{eqnarray}
For 221 degrees of freedom (2N minus three for $\tau$, $R$, and
$\bar{y}$), the probability of $\chi^2 \geq 301$ is 0.03\%.  The N=112
A light curve with the new model has PRH$\chi^2=114$, and the B light
curve has PRH$\chi^2=156$, thus the B image is the cause of the bad
fit, as we would expect since the Spring~1990 points are included.
The optimal reconstructions for N=112 A and B, using the new
covariance model, are comparable to Figures~\ref{N=107recon} and
\ref{N=107recondiff}, except the Spring~1990 points are more than five
$\sigma$ away from the reconstruction.

Finally, we analyzed the pseudo-jackknife data using PRH$Q$.  Delays
for the N=112 jackknife sets were scattered between 446 and 472~days
(-14 and +12~days from the delay found above for the real data).  This
scatter is comparable to the confidence interval from the Gaussian
Monte Carlo data.  For the N=107 jackknife data, where the five points
of Spring~1990 were already removed, the delays were scattered between
448 and 473~days (-12 and +13~days from the real data), except removal
of 1992~January~6 shifted the delay estimate by +21~days to 481~days.
This is not surprising, since 1992~January~6 is the first point after
the flux density decrease in the B image in 1991, and the alignment of this
feature with the A image decrease in 1990 is essential to the
calculation of the delay.  The removal of no single point caused the
time delay estimate to shift by more than twice the confidence interval
obtained from the Gaussian Monte Carlo data.  The scatter in time
delays obtained from the pseudo-jackknife data, which has all the
properties of the real data (including any non-Gaussian
characteristics), is confirmation that the confidence intervals
determined from the Gaussian Monte Carlo data are roughly correct.

\subsection{Comments} 

The time delay estimate found using PRH$\chi^2$ has changed
significantly since Press, Rybicki,~\&~Hewitt (PRHb) applied their
analysis to the first 80 data points in the VLA light curves (N=80).
The change is due to the new features that have entered the light
curves, particularly the flux density decrease in both images in
1990-91.  We reanalyzed the N=80 light curves, and the N=80 light
curves with five Spring~1990 points removed (N=75), trying both the
old and new covariance models.  With the old model, the N=80 and N=75
PRH$\chi^2$ surfaces have local minima (around 455 and 600~days), and
the global minimum in both cases is around 550~days.  With the new
covariance model, the N=80 and N=75 PRH$\chi^2$ surfaces are smooth,
with a single minimum at 540~days.  The minimum is also much broader
with the new covariance model, indicating a larger confidence
interval.  {\em Since removing points and changing the covariance
model has little effect on the best-fit delay in any version of the
light curves, the change in the delay estimate since 1992 is not due
to the new covariance model or to the removal of certain points, but
to the additional features.}  We confirm the finding of PRHa that the
choice of covariance model has little effect on the value of the best
fit time delay, but we find that it has a significant effect on the
topology of the PRH$\chi^2$ surface and the confidence interval.


\section{Dispersion Analysis}
\label{peltdisp}

To allow comparison with previous work in this area, we discuss in
this section and the next alternative analyses to the PRH$\chi^2$
method.  Here we evaluate the light curves using the dispersion method
of Pelt~{\it et~al.} (1994) and Pelt~{\it et~al.} (1996).

The dispersion method compares the flux densities of nearby pairs of
points and sums the squared differences over the whole curve.  To
measure the dispersion, the light curves from the A and B images are
first combined at a trial time delay $\tau$ and trial flux ratio
$R$. For this combined set of points, the dispersion is
\begin{equation}
	D^2(\tau,R) = \frac{ \sum_{ij} W_{ij}S_{ij}(a_i - b_j) ^2 }{ 2
\sum_{ij} W_{ij}S_{ij} },
\end{equation}
with weighting terms
\begin{equation}
S_{ij} = \left\{ \begin{array}{ll}
	1 - \frac{|t_i - t_i|}{\delta} & \mbox{if $|t_i-t_j| \leq \delta$} \\
	0                              & \mbox{if $|t_i-t_j| > \delta$}
		\end{array}
\right. 
\end{equation}
and 
\begin{equation}
	W_{ij} = \frac{W_iW_j}{W_i + W_j},
\end{equation}
where $W_i=1/e_i^2$ is the statistical weight for each observation. The
pairs included in this sum are all AB pairs in the combined curve with
separation less than $\delta=60$~days, which is of order 2N pairs.  The
$S_{ij}$ term, a modification to P94 added by P96, decreases the
weight on pairs with larger separations, and is essential for making
the dispersion a smooth function of time delay.  We used
$\delta=60\unit{days}$, just as in P96.  This weighting is in effect a type of
covariance model, where points less than 60~days apart are expected to
have identical flux densities, and points more than 60~days apart can
have any flux density difference.  The minimum of $D^2$ with respect to $\tau$
and R is the estimate of the time delay and flux ratio.

P94 also define a statistic $I$ to show the effects of the removal of
points,
\begin{equation}
I(l, m, \tau_1, \tau_2) = D^2(l, m, \tau_1) - D^2(l, m,
\tau_2),
\end{equation}
where $l$ is the location in the list of observations where $m$ points
are removed.  The statistic compares the dispersion at two different
delays ($\tau_1$ and $\tau_2$) when certain points are removed,
indicating whether those particular points favor one particular delay
or another.  P94 discussed mainly the case of $m=2$ points,
$\tau_1=536\unit{days}$, and $\tau_2=415\unit{days}$.

\begin{figure}
\plotfiddle{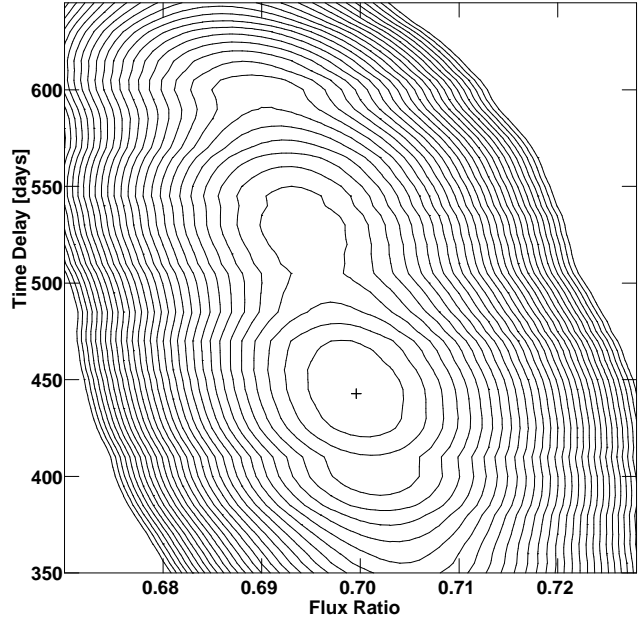}{3in}{0}{45}{45}{-135}{-75}
\caption{The dispersion surface for the N=112 light curves.  The 
global minimum is $D^2=0.01153$ at $\tau=443$, $R=0.6996$.  Contours
start at 0.0120 and increase by 0.0005 to 0.0260. }
\label{N=112dispsurf}
\end{figure}

\begin{figure}
\plotfiddle{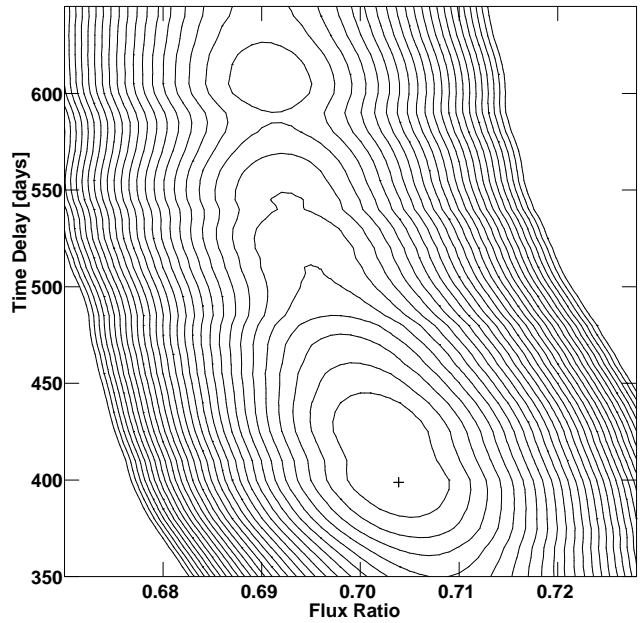}{3in}{0}{45}{45}{-135}{-75}
\caption{The dispersion surface for the N=107 light curves.  The 
global minimum is $D^2=0.00849$ at $\tau=399$, $R=0.7039$.  Contours
start at 0.0090 and increase by 0.0005 to 0.0230. }
\label{N=107dispsurf}
\end{figure}

Figures~\ref{N=112dispsurf} and \ref{N=107dispsurf} show the
dispersion surface in delay and ratio for the N=112 and N=107 data
sets, respectively.  In order to find the global minima of the
surfaces we again used a downhill simplex search (\cite{numrec}).  To
avoid local minima, we started ten ``amoeba'' searches at various
locations in the surface, then used the deepest point found as the
global minimum.  The global minimum for the N=112 real data set was
$D^2 = 0.01153\unit{dBJ^2}$ at $\tau=443$~days and $R=0.6996$.  The number of AB
pairs used in the calculation of the dispersion at the minimum was
276.  The bias and the 68\% confidence intervals were found from the
500 N=112 Gaussian Monte Carlo data sets, giving
\begin{eqnarray}
	\tau = 443^{+22}_{-21}\unit{days}, &R = 0.6995^{+0.0029}_{-0.0030}&
(N=112,\unit{dispersion}).
\end{eqnarray}
For N=107 points, the global minimum was $D^2 = 0.00849\unit{dBJ^2}$
at $\tau=399$~days and $R=0.7039$.  The number of pairs used in the
calculation at the minimum was 234.  There is also a secondary minimum
of $D^2 = 0.00854\unit{dBJ^2}$ at $\tau=427$~days and $R=0.7019$. The
bias and the 68\% confidence intervals were determined from the 500
N=107 Gaussian Monte Carlo data sets.  When the bias is taken into
account, the deepest minimum is
\begin{eqnarray}
	\tau =398^{+25}_{-22}\unit{days}, & R = 0.7039^{+0.0026}_{-0.0030}
&(N=107,\unit{dispersion}).  
\end{eqnarray}

P96 found that the value of the time delay for N=80, with no points
removed, was 616~days, and that the value shifted to 421~days when the
1990~April~10 and 1990~May~7 observations were removed.  When all five
Spring~1990 points are removed from the N=80 data set, the estimate of
the time delay shifts to 555~days.  Now with the new features in the
curve (N=112) the value of the delay is 443~days, and decreases to
398~days with the removal of the Spring~1990 points.

Delays for the N=112 pseudo-jackknife data, found using the
dispersion, were mostly around 442 to 444~days (just $\pm1$~day from
the delay found above for the real N=112 data), with only thirteen
data sets scattered out to 423 to 458~days (or -20 to +15~days from
the real data). This scatter is comparable to the confidence interval
found for the Monte Carlo data.  The N=107 pseudo-jackknife data had
delays at a few particular values, rather than a random scatter over a
range of delays: one data set at 456~days (1992~January~6, +57~days
from the real N=107 data), eighteen data sets at 427~days (+28~days
from the real data), four data sets at 401~days (+2~days from the real
data), and the rest at 399~days.  Since 1992~January~6 is the first
point after the flux density decrease in the B image in 1991, it is not
surprising that it is crucial to the calculation of the delay.  

This pseudo-jackknife test is not the same as the P94 $I$ statistic.
We only remove one point at a time, rather than two neighboring
points.  More importantly, our test determines the {\em best} delay
for each data set, while $I$ checks whether $\tau=415$~days or
$\tau=536$~days is a better fit (probably neither is the {\em best}
fit).  Finally, even if two points have a strong effect on the value
of the time delay, that does not mean they should be excluded from the
analysis, just that they are very influential to the final result (for
instance, if they occur during a rise or fall in the curve).  P94 and
P96 found that the removal of 1990~April~10 and 1990~May~7 shifted
the delay to a shorter value for the N=80 data set, and we confirm
that finding for the N=112 set (the delay estimate is shifted to 399
days). This in itself, however, is not a sufficient motivation to
exclude the points.


\section{Discrete Correlation Function Analysis}
\label{lndcf}

To be consistent with earlier work, we also analyzed the light curves
using a discrete correlation function.  We follow exactly the analysis
of L92, which is based on work by Edelson \&~Krolik (1988).

A cross-correlation function was one of the first statistical
techniques used to find the time delay (see Table~1).  The peak in the
correlation between two signals in time should be a reasonable
estimate of the delay between them.  However, serious errors can
result if the signal is irregularly sampled or if interpolation is
used (Falco~{\it et~al.}  1991b; PRHa; L92), so the correlation
function needs to be modified to handle a discrete set of points
rather than a continuously measured function.  The correlations must
found directly from discrete pairs of points in the A and B light
curves, and these correlations are then binned according to the time
separation between the A and B points.  So we use the Locally
Normalized Discrete Correlation Function (LNDCF)
\begin{equation}
	LNDCF(\tau) = \frac{1}{M}\sum_{ij}\frac{(a_i - \bar{a}_*)(b_j -
\bar{b}_*) }{[(\sigma^2_{a*} - e^2_a)(\sigma^2_{b*} - e^2_b)]^{1/2}},
\end{equation}
where the sum is over all AB pairs in the delay bin, {\it i.e.} all
pairs such that
\begin{equation}
	| t_i - t_j | - \tau \leq \frac{\Delta\tau}{2}.
\end{equation}
$M$ is the number of pairs in the bin, $\bar{x}_*$ and $\sigma_{x*}$
are the mean and standard deviation of the $x_i$ in the bin, $e_x$ are
measurement errors, and $\Delta\tau$ is the size of the bin around
$\tau$.  Since the mean and variance may not be constant for the
entire light curve ({\it i.e.} if the curve is not stationary), they
are calculated separately for each delay bin.  The LNDCF is binned by
definition, so it cannot be determined independently for all values of
$\tau$.  Decreasing the bin size improves the resolution with respect
to $\tau$, but also increases the error for each bin.  We used a bin
size of 30~days, the same as L92, and the number of pairs in this
calculation for the radio light curves is of order N/2.  To find the
time delay more precisely than the bin size, a cubic polynomial was
fit to the peak of the LNDCF.  An advantage of the LNDCF is that it is
independent of the flux ratio and only a function of time delay.  To
obtain the flux ratio, we combined the two curves using the fitted
delay, then adjusted the flux ratio to minimize the summed dispersion
between the curves.  The dispersion at each observed time was computed
using a linear interpolation of the adjacent points from the other
curve.

\begin{figure}
\plotone{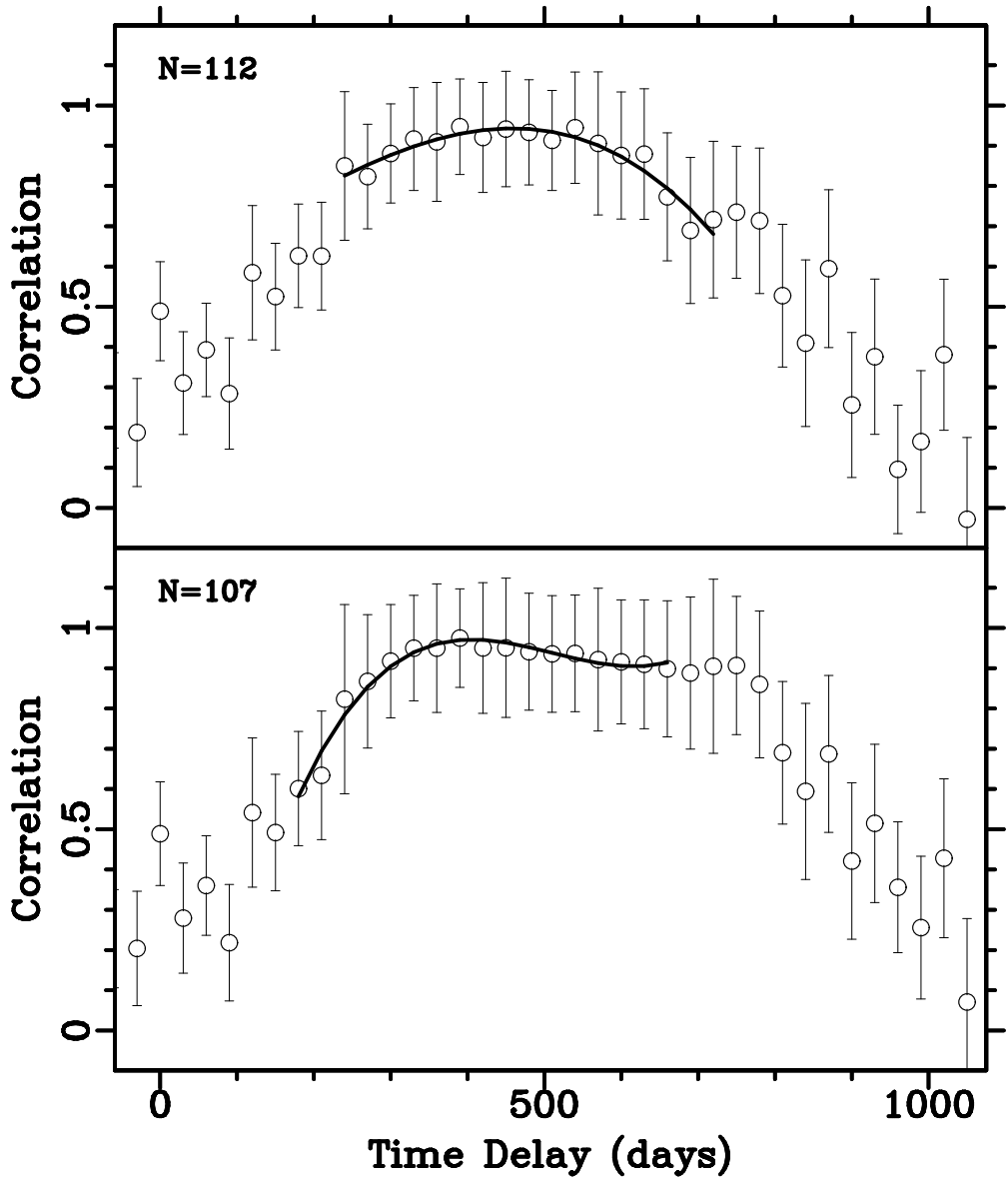}
\caption{The LNDCF (circles), and the cubic polynomial 
fit to the peak (thick line), for the N=112 and N=107 light curves. }
\label{N=112.107lndcf}
\end{figure}

The LNDCF for the N=112 and N=107 light curves is plotted in
Figure~\ref{N=112.107lndcf}, showing the LNDCF and its errors for each
delay bin, and the cubic fit to the peak of the LNDCF.  For N=112, the
maximum correlation of 0.943 is at $\tau=458$~days, where about 70
pairs were used in the calculation.  The fitted flux ratio for this
delay was $R=0.6980$.  The 68\% confidence intervals and the bias were
determined from the 500 N=112 Gaussian Monte Carlo data sets, giving
\begin{eqnarray}
\tau = 458^{+27}_{-27}\unit{days}, & R=0.6982^{+0.0027}_{-0.0026} & (N=112,\unit{LNDCF}).
\end{eqnarray}
For N=107, the maximum correlation of 0.971 was at $\tau=405$~days,
where about 60 pairs were used in the calculation.  The fitted flux
ratio for this delay was $R=0.7000$.  Using the 500 N=107 Gaussian
Monte Carlo data sets to find the bias and the 68\% confidence
intervals, 
\begin{eqnarray}
\tau = 404^{+25}_{-30}\unit{days}, & R = 0.6999^{+0.0030}_{-0.0026} & (N=107,\unit{LNDCF}).
\end{eqnarray}   

L92 found a value for the delay of $\tau=513$ (with a correlation of
0.97) when using this method on the first 80 data points.  With the
addition of the new features in the light curves, the estimate of the
delay has shifted to a lower value, in agreement with the PRH$Q$ and
dispersion results.  Unlike the PRH$Q$ result but in agreement with
the dispersion result, the delay estimate is significantly affected by
the removal of the five points in Spring~1990.  The LNDCF is dominated
by the flat portions of the light curve, which swamp out the signal
from sharp features, so the removal of the points from the strong 1990
rise make the LNDCF less able to distinguish between delays of
interest (note the flatness of the N=107 LNDCF from 350 to 600~days in
Figure~\ref{N=112.107lndcf}).

We calculated the LNDCF for the N=112 pseudo-jackknife data sets, and
found that the delays were scattered between 428 and 489~days (or -30
to +31~days from the delay found above for the real N=112 data),
except that removal of 1990~April~10 shifted the delay estimate
-44~days to 414~days. The scatter is comparable to the confidence
interval obtained from the Gaussian Monte Carlo data.  Removal of
1990~April~10 effectively selects the ``first rise'' in the B image in
1990 (see Section~7).  The LNDCF for the N=107 pseudo-jackknife data
found delays scattered between 395 and 417~days (-10 to +12~days from
the real N=107 data), with the removal of 1989~September~27 shifting
the delay estimate -19~days to 386~days.  The scatter is smaller than
the confidence interval obtained from the Monte Carlo data.  Since
1989~September~27 is the last point before the B rise in 1990, it is
crucial for alignment with the rise in 1988 of the A image.


\section{Discussion and Conclusions}
\label{conc}

We have calculated time delays by using three different statistical
techniques on the complete light curves, and on the light curves
with five points removed.  Confidence intervals have been found using
Gaussian process Monte Carlo data with the same sampling and structure
function as the real light curves.  Table~3 lists our results.  Note
that the flux ratio increases with decreasing time delay, a trend that
can be seen in the delay-ratio surfaces. Some time delay studies have
made the error of assuming one flux ratio while finding the delay,
rather than fitting for both parameters at once.  The degeneracy
between the fitted delay and ratio is probably due to the long decline
in the early part of the light curves.

Table~3 contains values ranging from 398 to 461~days.  Thus, the new
features in the light curves show convincingly that the time delay for
the radio light curves is less than 500~days.  But which of the delays
in Table~3 is correct?  We comment here on the advantages and
disadvantages of the different statistical methods used.

The PRH$Q$ statistic has the advantage of using all of the data in the
light curves ($4N^2$ pairs).  Each point is compared to every other
point, and the difference is checked for consistency with the
covariance model, rather than just comparing every point to its
nearest neighbors and checking if the difference is zero.  The method
is less dependent on the exact features of the light curve than the
other methods, and instead relies on the statistical properties of the
underlying quasar emission and the measurement error.  However,
PRH$Q$, as applied here, requires two important assumptions about the
data: that the light curve is statistically stationary (assumed when
we fit one structure function for the whole light curve), and that the
covariance model correctly describes the variations in the light
curve.  Since $Q$ was used to find the covariance model, the second
assumption should be a good one.  Although the use of $Q$ has reduced
the impact of sampling, the gaps in the observations still cause
delays of about 480~days to be excluded more strongly than other
delays for ``window function'' data.  We confirm the finding of PRHa
that the choice of covariance model for the PRH$\chi^2$ or $Q$
analysis has little affect on the best fit delay, but find that the
choice of covariance model can have a significant affect on the
topology of the PRH$\chi^2$ surface and the confidence interval found
from Monte Carlo analysis.

The dispersion method does not assume that the light curves are
stationary, but does assume that nearby points will have identical
flux densities if their separation is less than 60~days.  The
dispersion only uses about 2N pairs of points in the calculation.
While the original version of the dispersion given in P94 produced
very rough surfaces in time delay, the weighting modifications of P96
have made the dispersion a smooth function of delay and ratio, so that
there is a convincing (but broader) global minimum.

The discrete correlation function has the advantage of being
completely independent of the flux ratio, since it fits only in delay.
It assumes that the light curves are stationary within each delay bin,
and assumes that nearby points will have similar flux densities.  The
number of pairs going into a calculation is on the order of N/2.
Because of the bin size, the LNDCF has poor resolution in delay unless
it is interpolated at the peak.  Furthermore, it is dominated by the
non-performing parts of the signal, such as linear ramps and flat
sections, rather than the sharper features (thus, non-linear
variability is more important for this method than the others).

Given the arguments for and against each method, we are not convinced
that one method is superior to the others, and so we are not convinced
of a particular value of the time delay.  We {\em are} convinced that
the time delay, based on the current radio light curves, is in the
range 375 to 485~days, and definitely less than 500~days.  The flux
ratio is then in the range 0.695 to 0.707.  The large range for the
delay has two causes.  First, there have been unlucky coincidences of
key features in the light curves with gaps in the monitoring (the 1991
B decrease and 1994 A increase), or with unusual variations (the 1990
B increase).  Second, although all three methods agree on the time
delays inserted in stationary, Gaussian Monte Carlo light curves (with
$\sim$5\% uncertainty), they disagree by $\sim$15\% on the time delay
for the real light curves, indicating that the data may have
non-Gaussian properties.

Ideally, Monte Carlo data should be made to have {\em all} the
statistical properties of the real data. The only way to do this with
certainty is to derive the Monte Carlo light curves directly from the
real ones.  P94 and P96 did this by using a bootstrap technique to
make Monte Carlo data.  They first combined the light curves at their
best delay, then applied an adaptive median filter to the combined
curves, and finally bootstrapped the residuals.  Although they derive
the synthetic data from the real data, they also assumed a value for
the delay and and assumed that the residuals about the median-filtered
light curve are independent and identically distributed (probably not
the case).  Although our pseudo-jackknife analysis was not formally
correct, it was another way to produce synthetic light curves directly
from the data. We found that the scatter in the time delay estimates
for the pseudo-jackknife data was comparable to, and in some cases
better than, the confidence intervals from the Gaussian Monte Carlo
light curves.  To improve the statistical analysis of the current 6~cm
light curves beyond what has been done, the non-Gaussian
characteristics of this time series need to be incorporated into the
analysis.

\begin{figure}
\plotfiddle{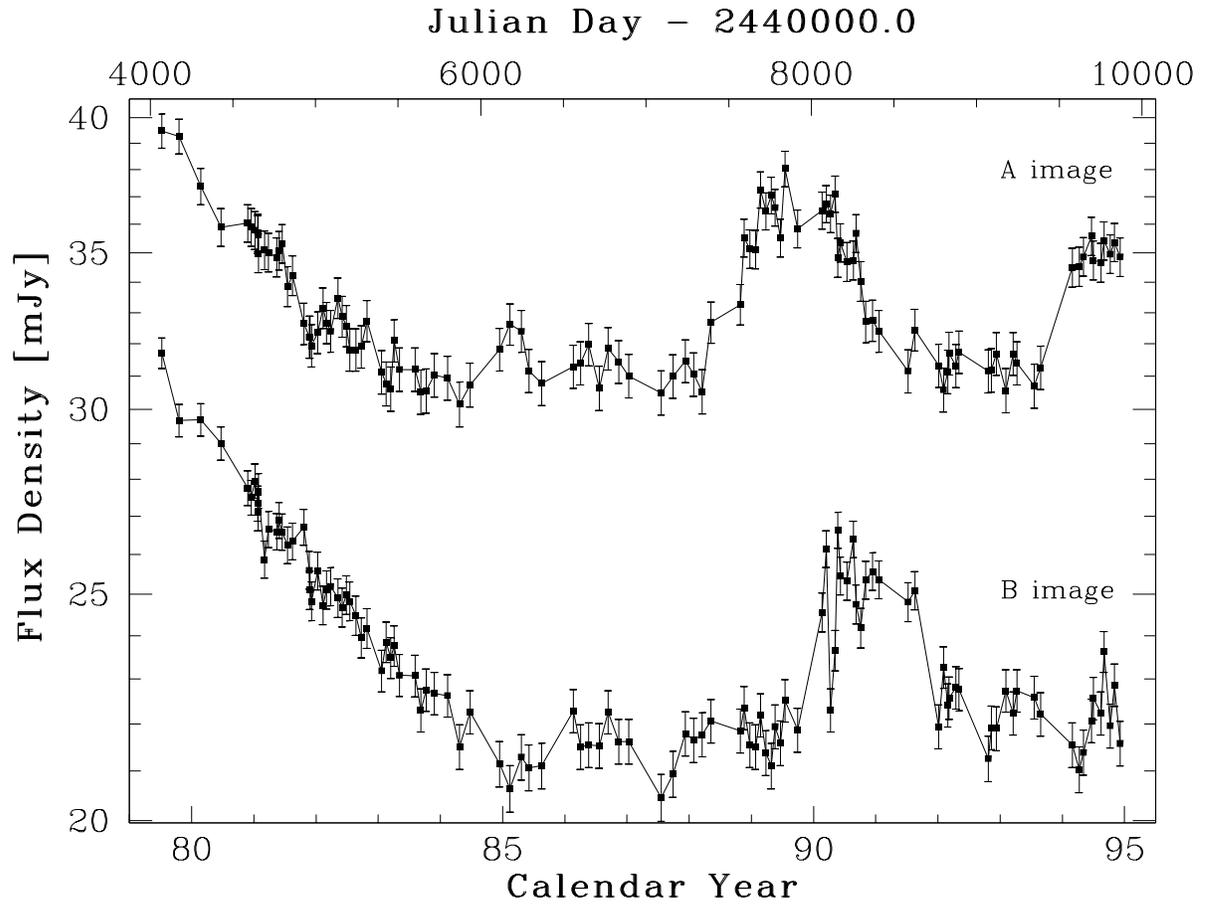}{6in}{-90}{60}{60}{-225}{475}
\caption{The 6~cm light curves of gravitational lens B0957+561, from 1979 to 1994.}
\label{lightcurve}
\end{figure}

\begin{figure}
\plotfiddle{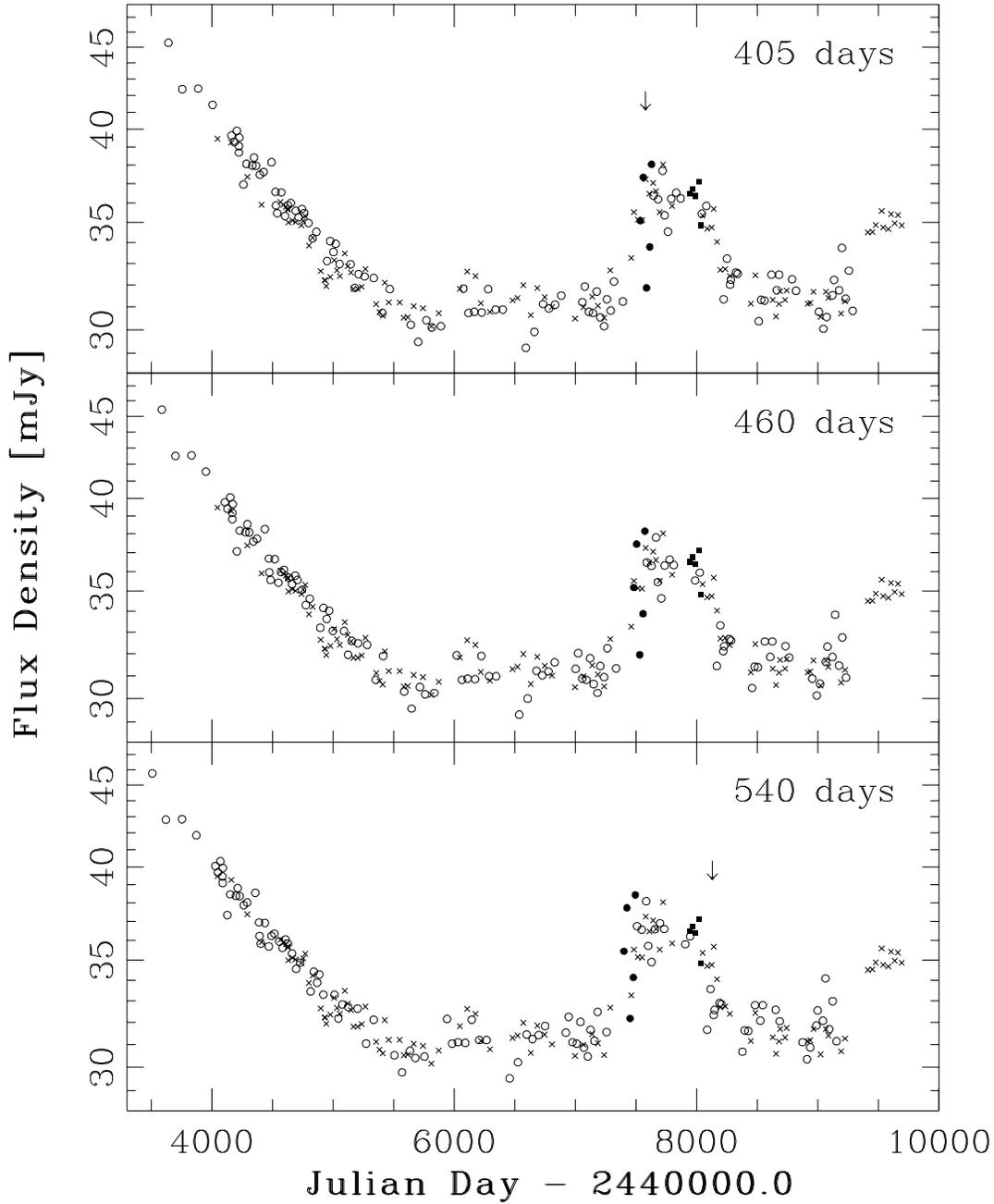}{6.5in}{0}{70}{70}{-225}{0}
\caption{Combined light curves.  The A light curve is shown 
with crosses, the B light curve is shown with open circles.  The five
points removed from A and B are filled squares and circles,
respectively.  The B light curve has been shifted back by
$\tau=405$\unit{days} and up by $R=0.700$ (top panel),
$\tau=460$\unit{days}, $R=0.698$ (middle panel), and
$\tau=540$\unit{days}, $R=0.693$ (bottom panel).  Error bars have been
omitted for clarity.  Arrows indicate epochs referred to in the text.}
\label{combine}
\end{figure}

The analyses discussed above are statistical, but often one can gain
insights into the problem by sliding the curves across one another to
get an estimate of the time delay by eye.  Figure~\ref{combine} shows
the light curves combined at three delays.  When combining the curves,
one finds that for longer delays (greater than 500~days), the A image
decline around Julian Day 2448000-8500 does not happen soon enough to
line up with the B image drop around Julian Day 2448500-2449000.  This
leads one to expect that the time delay we obtain based on the current
light curves will be shorter than the delay obtained previously, which
is exactly what we find from the statistical analysis above.  Note
that at shorter delays (less than 500~days) the fluctuation in the B
image in Spring~1990 starts to look out of place.  In fact, the B
image seems to have two possible rises, either in 1990~February-March
or in 1990~May, depending on which points you choose to ignore (see
Figure~\ref{lightcurve}).  Shorter time delays seem to correspond to
assuming the first rise is the real one, the longer time delays seem
to correspond to the second rise being real.  This explains why the
1990~April~10 and 1990~May~7 B image points are so influential to the
time delay analysis, as found by P94.  The PRH optimal reconstruction
gave us additional information about these points, indicating that the
observations in that epoch were produced by a different physical
process than the rest of the light curve.  Thus we removed the points
for both the first and the second rise in Spring~1990 for the N=107
analysis.  In the future we may gain a better understanding of the
different underlying physical processes, and may be able include those
in the model.

Since the A and B light curves show no large differences on time
scales of one to fifteen years, we conclude that the VLA 6~cm light
curves are not affected by microlensing or other mechanisms on these
time scales.  This is an advantage of the radio monitoring over
optical monitoring projects.  However, shorter fluctuations such as
that in the B image in Spring~1990 do confuse the analysis.  Such
variations may be due to microlensing or RISS of a compact jet
component emerging from the core of the quasar, and thus would be more
likely to occur during times of flux density increase than at other
times in the light curve.

An estimate of the time delay, when combined with a mass model, gives
us a measurement of the angular diameter distance to the lens.  If
assumptions are also made about the cosmology, then we can find the
Hubble parameter $H_0$.  While Kochanek (1991)
warns against modeling without sufficient observational constraints,
others make basic assumptions about the galaxy and cluster in the lens
and then do the fit.  Falco~{\it et~al.} (1991a) model the galaxy as a
King model sphere with a core mass, and the cluster as a shear effect.
Grogin~\&~Narayan (1996a, 1996b) model the galaxy as a softened
power-law sphere with a core radius, and the cluster as a shear.
Grogin~\&~Narayan (1996a, 1996b) fit both of these models to the VLBI
observations of Garrett~{\it et~al.}  (1994), assuming $\Omega=1$ and
$\Lambda=0$.  Grogin \&~Narayan found the fit to the King sphere lens
model to have a lower $\chi^2$ than the power-law sphere model,
although the reduced $\chi^2$ of 3.4 was still large, indicating a
poor fit to the observational constraints.  For the King sphere model
they find
\[
H_0 = 81.0^{+6.4}_{-6.1} \left(\frac{\sigma}{300\unit{km/s}}\right)^2 \left
(\frac{1.1\unit{years}}{\tau}\right)\unit{km\,s^{-1}Mpc^{-1}}.
\]
The velocity dispersion of the galaxy was measured recently by
Falco~{\it et~al.} (1996) to be
$\sigma = 259\pm10\unit{km/s}$ for positions more than 0.2\arcsec 
from the center of the lensing galaxy.  Using this velocity dispersion
and the shortest delay in Table~3, the King sphere model gives
$H_0=60.9^{+7.5}_{-7.6}$, while using the longest delay in Table~3
gives $H_0=52.6^{+6.1}_{-6.0}$.  The quoted uncertainties on these
values of $H_0$ are one $\sigma$, and 5\% error has been added for the
ambiguity due to large scale structure along the line of sight
(\cite{barkana96}).

Continued VLA monitoring of this source will improve our estimate of
the delay.  The quasar has been more variable in recent years than in
the 1980s, and new features may continue to enter the light curve.  We
have yet to achieve dense sampling of a sharp increase or decrease in
{\em both} the A and B light curves that is not confused by an unusual
variation.  Since 1990~October we have been monitoring B0957+561 at
4~cm as well as 6~cm.  At the shorter wavelength, the A and B images
are resolved in the VLA D array, so the 4~cm light curves will be free
of long gaps.  The flat-spectrum cores of the A and B images will be
better isolated from the extended emission in the image, and the cores
should also be more variable.  Finally, the multi-wavelength
information on the quasar variability will aid in understanding the
physical origin of the fluctuations.  Joint analysis of the 4~cm and
6~cm light curves will improve our measure of the time delay in the
0957+561 gravitational lens system, and will be reported in a future
publication.


\acknowledgements

We thank Dave Roberts, George Rybicki, Bill Press, Chris Moore, and
our referee for helpful comments.  This work was supported by a David
and Lucile Packard Fellowship in Science and Engineering, a NSF
Presidential Young Investigator Award, the MIT Class of 1948, and NSF
grants AST92-24191 and AST93-03527.


\clearpage

\begin{deluxetable}{lcccc}
\scriptsize
\tablecolumns{5}
\tablenum{1}
\tablewidth{7.25in}
\tablecaption{History of the Time Delay}
\tablehead{
 \colhead{Reference} &
 \colhead{Data Set} & 
 \colhead{Statistic} &
 \colhead{\parbox{.5in}{\begin{center}Delay (days)\end{center}}} &
 \colhead{\parbox{.5in}{\begin{center}Delay (years)\end{center}}}
}
\startdata
Lloyd 1981	             		& 28 optical obs 	& 1 & $>730$        	& $>2$  	\nl
					& over 2 years 		&   & 			& 		\nl
\tablevspace{6pt}
Keel 1982                 		& 17 optical obs 	& 1 & $>990$          	& $>2.7$        \nl
					& over 2 years  	&   & 			&               \nl
\tablevspace{6pt}
Florentin-Nielsen 1984			& 54 optical obs 	& 1 & $566\pm40$ 	& $1.55\pm0.1$  \nl
					& over 5 years 		&   & 			&               \nl
\tablevspace{6pt}
Bonometti 1985	        		& 3 VLBI obs 		& 1 & $470\pm260$    	& $1.3\pm0.7$ \nl
					& over 4 years   	&   & 			&               \nl
\tablevspace{6pt}
Gondhalekar {\it et al.} 1986  		& 11 UV obs 		& 1 & $660\pm70$     	& $1.8\pm0.2$ \nl
					& over 4 years   	&   & 			&               \nl
\tablevspace{6pt}
Schild \& Cholfin 1986      		& 28 optical obs 	& 3 & $376\pm40$     	& $1.03\pm0.1$  \nl
					& over 4 years   	&   & 			&               \nl
\tablevspace{6pt}
Leh\'ar {\it et al.} 1988		& 40 radio obs  	& 2 & $500\pm100$      	& $1.4\pm0.3$ \nl
					& over 8 years   	&   & 			&               \nl
\tablevspace{6pt}
Vanderriest {\it et al.} 1989 (V89)  	& 131 optical obs 	& 3, 4 & $415\pm20$ 	& $1.14\pm0.05$ \nl
					& over 8 years   	&   & 			&                \nl
\tablevspace{6pt}
Schild 1990 (S90)	      		& 329 optical obs  	& 3 & $404$             & $1.11$       \nl
					& over 10 years   	&   & 			&                \nl
\tablevspace{6pt}
Falco, Wambsganss, \& Schneider 1991 	& V89        		& 5 & $430\pm17$    	& $1.18\pm0.047$ \nl 
                              		& S90        		& 5 & $490\pm34$     	& $1.34\pm0.093$ \nl
\tablevspace{6pt}
Leh\'ar {\it et al.} 1992 (L92),     	& 80 radio obs 		& 6 & $513\pm40$        & $1.40\pm0.10$   \nl
Roberts {\it et al.} 1991		& over 11 years   	&   & 			&                \nl
\tablevspace{6pt}
Press {\it et al.} 1992a          	& V89        		& 7 & $536^{+14}_{-12}$ & $1.47^{+0.038}_{-0.033}$ \nl
\tablevspace{6pt}
Press {\it et al.} 1992b          	& L92        		& 7 & $548^{+19}_{-16}$ & $1.50^{+0.052}_{-0.044}$ \nl
                              		& V89, L92   		& 7 & $540\pm12$        & $1.48\pm0.033$  \nl
\tablevspace{6pt}
Oknyanskij \& Beskin 1993   		& L92        		& 8 & $540\pm30$        & $1.48\pm0.082$  \nl
\tablevspace{6pt}
Pelt {\it et al.} 1994            	& V89, S90   		& 9 & $415\pm32$       & $1.14\pm0.088$  \nl
                              		& L92*        		& 9 & $409\pm23$       & $1.12\pm0.063$  \nl
\tablevspace{6pt}
Beskin \& Oknyanskij 1995   		& V89, S90   		& 8 & $530\pm15$      	& $1.45\pm0.04$     \nl
\tablevspace{6pt}
Schild \& Thomson 1995 (ST95)		& 835 optical obs	& \nodata  & \nodata  	& \nodata	     \nl
					& over 16 years   	&   & 			&                   \nl
\tablevspace{6pt}
Campbell {\it et al.} 1995 		& 4 VLBI obs 		& 1 & $\sim365$		& $\sim1$	\nl
					& over 6 years		&   &			&		\nl
\tablevspace{6pt}
Pelt {\it et al.} 1996            	& ST95       		& 10 & $423\pm6$       	& $1.16\pm0.016$   \nl
			      		& L92*        		& 10 & $421\pm25$       & $1.15\pm0.068$   \nl
\tablevspace{6pt}
Kundi\'c {\it et al.} 1995, 1996	& $\sim$100 optical obs	& 3, 7, 10 & $418.5\pm6.0$ & $1.146\pm0.016$   \nl
					& over 2 years		&    &			&		\nl
\tablevspace{6pt}	
\tablebreak
Oscoz {\it et al.} 1996, Kundi\'c {\it et al.} 1995	& 40 optical obs	& 1  & $<500$		& $<1.4$	\nl
					& over 1 year		&    &			&		\nl
\tablevspace{6pt}
Thomson \& Schild 1997 			& ST95 			& 11 & $405\pm12$       & $1.11\pm0.033$   \nl
					& 		  	&   & 			&                  \nl
\enddata
\tablecomments{Time delays were converted from number given in reference using 1 year = 365.25 days.
Statistical techniques are: 
1) Inspection, 
2) Polynomial Fitting, 
3) Interpolation and Cross Correlation, 
4) Discrete Fourier Analysis,
5) Dispersion (all pairs), 
6) Locally Normalized Discrete Correlation,
7) $\chi^2$ Structure Function Analysis, 
8) Flux Ratio Dispersion,
9) Dispersion (near-neighbors), 
10) Weighted Dispersion (near-neighbors), and
11) Interpolation, Filtering, and Cross Correlation.
*1990~April~10 and 1990~May~7 removed }
\end{deluxetable}


\clearpage

\begin{deluxetable}{cccccc}
\scriptsize
\tablecolumns{6}
\tablewidth{0pt}
\tablenum{2}
\tablecaption{VLA 6 cm Light Curve Data}

\tablehead{
 \colhead{\parbox{.5in}{\begin{center}Calendar Date\end{center}}} &
 \colhead{\parbox{.5in}{\begin{center}Observation Array\end{center}}} &
 \colhead{\parbox{.5in}{\begin{center}Julian Day minus 2440000.0\end{center}}}&
 \colhead{\parbox{.5in}{\begin{center}A flux density (mJy)\end{center}}} &
 \colhead{\parbox{.5in}{\begin{center}B flux density (mJy)\end{center}}} &
 \colhead{\parbox{.5in}{\begin{center}Reduction Array\end{center}}} 
}

\startdata
79Jun23  & P    & 4047.5\phn    & 39.47  & 31.71 & A  \nl
79Oct13  & P    & 4160.16   & 39.26  & 29.67 & A  \nl
80Feb23  & P    & 4292.79   & 37.37  & 29.69 & A  \nl
80Jun20  & P    & 4411.30   & 35.90  & 29.01 & A  \nl
80Nov24  & A    & 4567.93   & 36.04  & 27.76 & A  \nl
80Dec16  & A    & 4589.5\phn    & 35.90  & 27.50 & A  \nl
81Jan06  & A    & 4610.99   & 35.79  & 27.95 & A  \nl
81Jan23  & A    & 4628.25   & 35.64  & 27.34 & A  \nl
81Jan24  & A    & 4628.75   & 35.70  & 27.09 & A  \nl
81Jan26  & A    & 4631.47   & 34.99  & 27.68 & A  \nl
81Mar03  & A    & 4666.71   & 35.10  & 25.87 & A  \nl
81Mar27  & A    & 4690.71   & 35.02  & 26.65 & A  \nl
81May14  & B    & 4738.61   & 34.85  & 26.59 & B  \nl
81May28  & B    & 4752.68   & 35.08  & 26.89 & B  \nl
81Jun14  & B    & 4769.55   & 35.32  & 26.58 & B  \nl
81Jul16  & B    & 4802.41   & 33.86  & 26.24 & B  \nl
81Aug15  & B    & 4832.30   & 34.23  & 26.34 & B  \nl
81Oct20  & C    & 4898.16   & 32.64  & 26.71 & C  \nl
81Nov21  & C    & 4930.08   & 32.22  & 25.60 & C  \nl
81Nov25  & C    & 4934.00   & 32.22  & 25.10 & C  \nl
81Dec05  & C    & 4944.06   & 31.94  & 24.83 & C  \nl
82Jan09  & C    & 4978.97   & 32.36  & 25.58 & C  \nl
82Feb09  & A    & 5009.81   & 33.15  & 24.73 & A  \nl
82Mar03  & A    & 5031.75   & 32.67  & 25.11 & A  \nl
82Mar27  & A    & 5055.64   & 32.40  & 25.19 & A  \nl
82May08  & A    & 5097.61   & 33.48  & 24.91 & A  \nl
82Jun03  & A    & 5123.52   & 32.87  & 24.68 & A  \nl
82Jun27  & A    & 5148.41   & 32.57  & 24.98 & A  \nl
82Jul16  & B    & 5167.32   & 31.81  & 24.83 & B  \nl
82Aug21  & B    & 5203.30   & 31.81  & 24.48 & B  \nl
82Sep23  & B    & 5236.13   & 31.91  & 23.95 & B  \nl
82Oct25  & B$\rightarrow$D & 5268.14   & 32.73  & 24.17 & B \nl   
83Jan20  & C    & 5354.89   & 31.12  & 23.18 & C  \nl
83Feb16  & C    & 5381.81   & 30.77  & 23.85 & C  \nl
83Mar15  & C    & 5408.82   & 30.61  & 23.48 & C  \nl
83Apr03  & C    & 5427.67   & 32.10  & 23.76 & C  \nl
83May05  & C    & 5459.62   & 31.20  & 23.08 & C  \nl
83Aug04  & A    & 5551.41   & 31.20  & 23.07 & A  \nl
83Sep06  & A    & 5584.30   & 30.52  & 22.30 & A  \nl
83Oct08  & A    & 5616.19   & 30.55  & 22.75 & A  \nl
83Nov26  & A    & 5665.06   & 31.03  & 22.67 & A  \nl
84Feb11  & B    & 5741.85   & 30.94  & 22.62 & B  \nl
84Apr22  & C    & 5812.69   & 30.15  & 21.51 & C  \nl
84Jun22  & C    & 5874.43   & 30.73  & 22.26 & C  \nl
84Dec12  & A    & 6046.95   & 31.82  & 21.15 & A  \nl
85Feb12  & A    & 6108.86   & 32.62  & 20.64 & A  \nl
85Apr20  & B    & 6175.72   & 32.40  & 21.29 & B  \nl
85Jun03  & B    & 6219.55   & 31.16  & 21.07 & B  \nl
85Aug17  & C    & 6295.25   & 30.78  & 21.11 & C  \nl
86Feb19  & A    & 6480.80   & 31.29  & 22.28 & A  \nl
86Apr03  & A    & 6523.66   & 31.40  & 21.51 & A  \nl
86May21  & A    & 6571.59   & 31.98  & 21.55 & A  \nl
86Jul20  & B    & 6632.28   & 30.63  & 21.53 & B  \nl
86Sep11  & B    & 6685.12   & 31.85  & 22.26 & B  \nl
86Nov12  & C    & 6747.11   & 31.44  & 21.62 & C  \nl
87Jan11  & C    & 6806.89   & 31.00  & 21.62 & C  \nl
87Jul20  & A    & 6997.35   & 30.49  & 20.46 & A  \nl
87Sep27  & A    & 7066.27   & 30.99  & 20.94 & A  \nl
87Dec09  & B    & 7138.99   & 31.46  & 21.79 & B  \nl
88Jan26  & B    & 7186.83   & 31.05  & 21.65 & B  \nl
88Mar17  & C\&D  & 7237.71   & 30.53  & 21.77 & C  \nl
88May08  & C\&D  & 7289.59   & 32.68  & 22.06 & C  \nl
88Oct27  & A    & 7462.12   & 33.26  & 21.85 & A  \nl
88Nov18  & A    & 7484.03   & 35.52  & 22.35 & A  \nl
88Dec21  & A    & 7516.98   & 35.14  & 21.55 & A  \nl
89Jan24  & A    & 7550.94   & 35.12  & 21.51 & A  \nl
89Feb24  & AnB  & 7581.75   & 37.24  & 22.19 & A  \nl
89Mar25  & B    & 7610.71   & 36.47  & 21.38 & B  \nl
89Apr26  & B    & 7642.65   & 37.04  & 21.11 & B  \nl
89May19  & BnC  & 7665.56   & 36.60  & 21.94 & B  \nl
89Jun20  & C    & 7697.59   & 35.52  & 21.59 & C  \nl
89Jul16  & C    & 7724.40   & 38.03  & 22.51 & C  \nl
89Sep27  & C    & 7797.28   & 35.84  & 21.87 & C  \nl
90Feb19  & A    & 7941.84   & 36.49  & 24.56 & A  \nl
90Mar15  & A    & 7965.82   & 36.73  & 26.14 & A  \nl
90Apr10  & A    & 7991.65   & 36.38  & 22.30 & A  \nl
90May07  & A$\rightarrow$AnB & 8018.72   & 37.09  & 23.65 & A  \nl 
90May23  & A$\rightarrow$AnB & 8034.53   & 34.84  & 26.63 & A  \nl
90Jun07  & A$\rightarrow$AnB & 8049.57   & 35.35  & 25.46 & A  \nl
90Jul15  & AnB    & 8088.36   & 34.69  & 25.33 & A  \nl
90Aug21  & B    & 8125.32   & 34.75  & 26.39 & B  \nl
90Sep06  & B    & 8141.21   & 35.68  & 24.75 & B  \nl
90Oct04  & BnC  & 8169.22   & 34.04  & 24.18 & B  \nl
90Nov01  & C    & 8197.06   & 32.70  & 25.35 & C  \nl
90Dec13  & C    & 8238.89   & 32.74  & 25.57 & C  \nl
91Jan17  & C    & 8273.79   & 32.40  & 25.36 & C  \nl
91Jul10  & A    & 8448.40   & 31.15  & 24.81 & A  \nl
91Aug18  & A    & 8487.31   & 32.44  & 25.09 & A  \nl
92Jan06  & B    & 8627.97   & 31.32  & 21.94 & B  \nl
92Feb04  & BnC  & 8656.80   & 30.58  & 23.26 & B  \nl
92Feb29  & C    & 8681.74   & 31.13  & 22.41 & C  \nl
92Mar07  & C    & 8688.67   & 31.70  & 22.57 & C  \nl
92Apr18  & C    & 8730.60   & 31.31  & 22.80 & C  \nl
92May03  & C    & 8745.60   & 31.74  & 22.76 & C  \nl
92Oct23  & A    & 8919.08   & 31.15  & 21.26 & A  \nl
92Nov11  & A    & 8938.09   & 31.18  & 21.92 & A  \nl
92Dec10  & A    & 8966.97   & 31.69  & 21.90 & A  \nl
93Feb05  & AnB  & 9023.78   & 30.56  & 22.73 & A  \nl
93Mar21  & B    & 9067.64   & 31.69  & 22.23 & B  \nl
93Apr09  & B    & 9086.67   & 31.40  & 22.73 & B  \nl
93Jul25  & C    & 9194.21   & 30.70  & 22.58 & C  \nl
93Aug26  & C    & 9226.26   & 31.26  & 22.21 & C  \nl
94Mar04  & A    & 9415.73   & 34.50  & 21.55 & A  \nl
94Apr11  & A    & 9453.68   & 34.53  & 21.03 & A  \nl
94May07  & A$\rightarrow$AnB & 9479.63   & 34.87   & 21.39  & A  \nl   
94Jun25  & B    & 9528.52   & 35.58  & 22.07 & B  \nl
94Jul06  & B    & 9540.42   & 34.75  & 22.56 & B  \nl
94Aug18  & B    & 9583.28   & 34.67  & 22.23 & B  \nl
94Sep08  & B    & 9604.27   & 35.42  & 23.62 & B  \nl
94Oct10  & BnC  & 9636.18   & 34.96  & 21.96  & B \nl
94Nov07  & C    & 9664.08   & 35.37  & 22.86 & C  \nl
94Dec08  & C    & 9694.92   & 34.86  & 21.58 & C  \nl
\enddata
\end{deluxetable}


\clearpage

\begin{deluxetable}{lccc}
\tablecolumns{4}
\tablenum{3}
\tablewidth{0pt}
\tablecaption{Results}
\tablehead{
 \colhead{Statistic} &
 \colhead{Light Curve} & 
 \colhead{\parbox{.5in}{\begin{center}Time Delay (days)\end{center}}} &
 \colhead{Flux Ratio} 
}
\startdata
PRH$Q$       & N=112 & $459^{+14}_{-16}$ & $0.6976^{+0.0022}_{-0.0023}$ \nl
             & N=107 & $461^{+16}_{-15}$ & $0.6981^{+0.0023}_{-0.0024}$ \nl
Dispersion   & N=112 & $443^{+22}_{-21}$ & $0.6995^{+0.0029}_{-0.0030}$ \nl
             & N=107 & $398^{+25}_{-22}$ & $0.7039^{+0.0026}_{-0.0030}$ \nl
LNDCF        & N=112 & $458^{+27}_{-27}$ & $0.6982^{+0.0027}_{-0.0026}$ \nl
             & N=107 & $404^{+25}_{-30}$ & $0.6999^{+0.0030}_{-0.0026}$ \nl
\enddata
\end{deluxetable}



\clearpage

\end{document}